\documentclass[prb,superscriptaddress,longbibliography,twocolumn]{revtex4-2}

\usepackage{graphicx}
\usepackage{hyperref}
\usepackage{amsmath}
\usepackage{amssymb}
\usepackage{bm}
\usepackage{units}
\usepackage{xcolor}
\usepackage{ulem}

\usepackage{amsfonts}
\usepackage{tabularx, ragged2e}
\usepackage{tcolorbox}
\usepackage{appendix}
\usepackage{mathptmx}
\usepackage{subfiles}
\usepackage{multirow}
\usepackage{mathtools}
\usepackage{xltabular}
\usepackage{ragged2e}
\newcolumntype{Y}{>{\RaggedRight\arraybackslash}X}
\newcolumntype{W}{>{\Centering\arraybackslash}X}
\newcommand{\TO}{\begin{tcolorbox}[width=\textwidth,colback={green}]}
	\newcommand{\DO}{\end{tcolorbox}}
\definecolor{muschi}{cmyk}{0,.2,0,0} 
\newcommand{\AT}{\begin{tcolorbox}[width=\textwidth,colback={muschi}]}
	\newcommand{\TA}{\end{tcolorbox}}
\definecolor{blut}{cmyk}{0,.4,0,0} 
\newcommand{\PRO}{\begin{tcolorbox}[width=\textwidth,colback={blut}]}
	\newcommand{\BLEM}{\end{tcolorbox}}
\definecolor{todo}{cmyk}{0,0,0,.45}

\definecolor{boldi}{cmyk}{.9,.1,0,.55} 

\definecolor{boldy}{cmyk}{0,.9,.5,.25}

\newcommand{\lcpq}{Univ Toulouse, CNRS, LCPQ, Toulouse, France}
\newcommand{\lpt}{Univ Toulouse, CNRS, LPT, Toulouse, France}
\newcommand{\etsf}{European Theoretical Spectroscopy Facility (ETSF), Toulouse, France}

\begin{document}
	
	\title{Core and valence photoemission spectra of atoms and molecules from a multichannel Dyson equation}
	
	\author{Stefano Paggi}
 \email{spaggi@irsamc.ups-tlse.fr}
	\affiliation{\lpt}
	\affiliation{\etsf}
		\author{J. Arjan Berger}
	\email{arjan.berger@irsamc.ups-tlse.fr}
	\affiliation{\lcpq}
	\affiliation{\etsf}
 \author{Pina Romaniello}
 \email{pina.romaniello@irsamc.ups-tlse.fr}
	\affiliation{\lpt}
	\affiliation{\etsf}
	
\begin{abstract}
We recently presented multichannel Dyson equations for the \textit{ab initio} simulation of various spectroscopies.
In particular, we introduced a multichannel Dyson equation for the description of photoemission spectra.
In this work, we apply our approach to the simulation of photoemission spectra of atoms and molecules.
We introduce a numerically efficient approach to calculate their spectral functions.
We compare the spectra obtained within the multichannel Dyson equation to those obtained with full configuration interaction and the $GW$ method.
We are thus able to show that the satellite features due to shake-up processes are significantly better described by the multichannel Dyson equation than by $GW$.
Finally, we also discuss the slow convergence of the satellite energies with the size of the basis set and we propose a simple extrapolation method to reach the complete basis-set limit.
\end{abstract}
	
	\maketitle

\section{Introduction}
Ultraviolet and X-ray photoemission spectroscopy (UPS and XPS) are powerful experimental techniques used to probe the electronic structure of atoms and molecules by measuring the kinetic energy of electrons ejected upon irradiation with high-energy photons~\cite{hufner}. While UPS primarily accesses valence electrons and is widely used to study chemical bonding and molecular orbitals~\cite{Fischer_2022}, XPS probes core-level electrons, providing element-specific information and insight into oxidation states and local chemical environments~\cite{deGroot_2001,deGroot_2008,vanderHeide_2011,Stevie_2020}. These methods are invaluable in chemistry because they give direct experimental access to ionization energies and electronic structure, helping to test and refine theoretical methods. A key feature of photoemission spectra is the presence of satellite structures, which are additional peaks beyond the main ionization lines, that arise from electron correlation effects. In particular, shake-up processes, where ionization is accompanied by the excitation of another electron, produce distinct satellites that encode detailed information about many-body interactions and electronic relaxation, making them essential for a deeper and more accurate interpretation of complex systems\cite{deGroot_2001,deGroot_2008,vanderHeide_2011,Stevie_2020,Bagus_2018}.

From a theoretical perspective, UPS and XPS spectra are most often simulated using Green's function approaches since the poles of the 1-body Green's function correspond to the electron removal energies 
which includes both quasi-particle and satellite energies.
The most common Green's function approach to simulate UPS and XPS spectra is the $GW$ method~\cite{Hed65,Hybertsen_1986,Ary98,Hed99,Reining_2018,Golze_2019}.
While $GW$ can accurately predict quasi-particle energies~\cite{Hybertsen_1986,Ary98,Reining_2018,Golze_2019,Ary98,VanS06,GW100,Li_2022}, it fails to correctly describe satellite energies.
In particular, $GW$ greatly overestimates the binding energies of the satellites~\cite{Lan70,Ary96,Guz11,riva_prl,Mejuto_2021}.
We note that the $GW$ method also suffers from other shortcomings~\cite{Lan12,Ber14,Sta15,Loo18,Vér18,Tar17}.
Combining the $GW$ method with the cumulant expansion approximation generally improves the description of the satellites~\cite{Ary96,Guz11,Lischner_2013,Zhou_2015,Gumhalter_2016}.
However, when applied to atoms and molecules it introduces unphysical excitations~\cite{Kocklauner_2025}.
Wave-function based methods such as configuration interaction and coupled-cluster theory can also produce accurate quasi-particle and satellite energies
~\cite{Brisk_1975,Arneberg_1982,Chatterjee_2019,Nakatsuji_1991,Kuramoto_2004,Marie_2024,Lisini_1992,Lisini_1993,Fronzoni_1999}.
Finally, wave-function theories can be used to obtain the one-body Green’s function through its definition in terms of many-body wave functions and energies, thus allowing for the calculation photoemission spectra~\cite{Nooijen_1992,Nooijen_1993,Nooijen_1995,Rehr_2020,Vila_2020,Vila_2021,Vila_2022}.
However, all these wave-function approaches are numerically more expensive than $GW$ and are, therefore, limited to small molecules.

In recent years, we have developed the multichannel Dyson equation (MCDE) in which two or more many-body independent-particle Green's functions are coupled through a multichannel self-energy~\cite{riva_prl,riva_prb,riva_prb_25,paggi_25,Sellie_2026}.
In particular, for photoemission spectroscopy, we introduced the (3,1)-MCDE that couples the Hartree-Fock one-body-Green's function (1-GF) to the 2-hole-1-electron ($2h1e$) and 2-electron-1-hole ($2e1h$) channels of the Hartree-Fock three-body Green's function (3-GF)\cite{Riv22}.
While the 1-GF contains information about quasi-particles the $2h1e$ and $2e1h$ channels of the Hartree-Fock 3-GF contain information about satellites, thereby putting quasiparticles and satellites on equal footing.
We have recently applied our approach to the \textit{ab initio} simulation of the direct and inverse photoemission spectrum of bulk silicon~\cite{Romaniello_2026}.
We have also shown the connection between the algebraic diagrammatic construction (ADC) method~\cite{Sch83,Nie84} and the MCDE~\cite{Demartini_2026}. 
In particular, we have shown that the $(3,1)$-MCDE is equivalent to the 2-particle-hole-Tamm-Dancoff approximation (2ph-TDA)~\cite{Sch78} (also known as ADC(2)-X).
However, higher order MCDE approximations are different from the ADC approximations~\cite{riva_prb_25}.
Another important advantage of the MCDE is that the it provides a physically more transparent framework than ADC, thus making it simpler to propose physically motivated improvements~\cite{riva_prb}.
Finally, we have recently performed a detailed comparison of the performance of MCDE and ADC methods for the calculation of quasiparticle energies of molecules~\cite{Keizer_2026}.

In this work we extend the MCDE to the simulation of the UPS and XPS spectra of atoms and molecules 
and we will assess its quality by comparing to spectra obtained using full configuration interaction (FCI) and $GW$.
We will also describe a numerically efficient approach to obtain the spectral function from the MCDE.
Finally, we will demonstrate the extremely slow convergence with the size of the basis set of the satellites in the spectral functions.
To solve this problem, we propose a simple extrapolation method to reach the complete basis set limit for both quasiparticle and satellite energies.

The article is organized as folllows.
In section \ref{sec:theory} we give a reminder of the MCDE equations one has to solve in practice. 
Then, in section \ref{sec:implementation}, we discuss the details of our implementation and in section \ref{sec:compdetails} we give some computational details.
In section \ref{sec:results} we report the spectral functions for atoms and molecules we obtained with the multichannel Dyson equation and compare them to FCI and $GW$ spectra.
We also show how to obtain MCDE spectral functions in the complet basis set limit.
Finally, in section \ref{sec:conclusions} we draw conclusions from our work and give an outlook on future developments.
\section{Theory}
\label{sec:theory}
To describe photoemission spectra, we will use the $(3,1)$-MCDE which describes the removal or addition of an electron 
by coupling the Hartree-Fock 1-GF to the $2h1e$ and $2e1h$ channels of the Hartree-Fock 3-GF~\cite{riva_prl,riva_prb}.
It is convenient to rewrite the $(3,1)$-MCDE as an eigenvalue problem of an effective Hamiltonian.
The latter is given by
\begin{equation}\label{Heff:eq}
	H_3^{\text{eff}}=\begin{pmatrix}
		H^{1\text p} & H^{\text{1p/3p}}  \\
		H^{\text{3p/1p}}  & H^{3\text p}
	\end{pmatrix},
\end{equation}
in which
\begin{align}
	H^{1\text p}_{i;m}&=  \epsilon_i \delta_{im},
	\\
	H^{\text{1p/3p}}_{i;mok}&= (f_{m}-f_{k}) (f_{o}-f_{k}) \Sigma^{\text{1p/3p}}_{i;mok},
	\\
	H^{\text{3p/1p}}_{ijl;m} &=  (f_{i}-f_{l})(f_{j}-f_{l}) \Sigma^{\text{3p/1p}}_{ijl;m},
	\\
	H^{3\text p}_{i>jl;m>ok}&= (\epsilon_{i}+\epsilon_{j}-\epsilon_{l}) \delta_{im}\delta_{jo}\delta_{lk}\nonumber\\
	&+ (f_{i}-f_{l}) (f_{j}-f_{l}) \Sigma_{ijl;mok}^{3p}.\label{eq:h3p}
\end{align}
where $f_i$ and $\epsilon_i$ are the occupation number and energy, respectively, of the Hartree-Fock spin-orbital $\varphi_i$.
At $T=0$ K, which is the case treated here, the occupation numbers are 1 for occupied orbitals and or 0 for unoccupied orbitals.
The conditions on the spin-orbitals $i>j$ and $m>o$ in Eq.~\eqref{eq:h3p} prevent the double counting of contributions.
The various contributions to the the multichannel self-energy are given by 
\begin{align}
	\Sigma^{3 \text p}_{ijl;mok}&=\!\! [(1\!-\!f_i)\!(1\!-\!f_j)f_l\!-\!f_if_j(1\!-\!f_l)][\delta_{lk} \bar v_{ijom} \nonumber \\ &+\!\delta_{mj}\bar v_{iklo} \!+\! \delta_{io} \bar v_{jklm} \!-\! \delta_{oj}  \bar v_{iklm} \!-\! \delta_{im}  \bar v_{jklo}] ,\label{selfthird:eq}\\
	\Sigma^{\text{1p/3p}}_{i;mok}&=\bar v_{ikom}, \label{selfcoupling:eq} \\
	\Sigma^{\text{3p/1p}}_{ijl;m}&=\bar v_{ijlm}, \label{selfcouplingtilde:eq}
\end{align}
where $\bar{v}_{ijlm}=v_{ijlm}-v_{ijml}$ with the matrix element of the Coulomb interaction $v(\vec{r}_1, \vec{r}_2)$ given by
\begin{equation}
v_{ijlm}=\iint dx_1 dx_2  \varphi^\ast_i(x_1)\varphi^\ast_j(x_2)v(\vec{r}_1, \vec{r}_2)\varphi_l(x_2)\varphi_m(x_1).
\end{equation}
where $x$ refers to both position and spin variables.
The occupation numbers in Eq.~\eqref{selfthird:eq} ensure that the hole-hole-electron and electron-electron-hole channels do not mix in $\Sigma^{3 \text p}$.
While $\Sigma^{3 \text p}$ dresses the three independent particles (2 holes and 1 electron or 1 hole and 2 electrons) in the 3-particle channel, $\Sigma^{\text{1p/3p}}$ and $\Sigma^{\text{3p/1p}}$ couple the 3-particle channel to the 1-particle channel,
thus dressing the independent particles in the 1-particle channel.
We note that, within this approximation to the multichannel self-energy, the effective Hamiltonian in Eq.~\eqref{Heff:eq} is equivalent to the 2ph-TDA method~\cite{Sch78} which is also known as ADC(2)-X.
For a detailed discussion on the link between the (3,1)-MCDE equations and the ADC equations we refer the reader to Ref.~\cite{Demartini_2026}.
Finally, we note that we have also recently performed a benchmark of MCDE and ADC methods for the quasiparticle energies of small molecules~\cite{Keizer_2026}.

From the eigenvalues $E_{\lambda}$ and eigenvectors  $A_{\lambda}$ of the effective Hamiltonian $H_3^{\text{eff}}$ we can construct the one-body Green's function according to
\begin{equation}
\label{Eqn:G1_MCDE}
    G_{i;m}(\omega) =  \sum_{\lambda}\frac{A^i_{\lambda}A^{*m}_{\lambda}}{\omega-E_{\lambda}+i\eta\text{sgn}(E_{\lambda}-\mu)}
\end{equation}
in which $\mu$ is the chemical potential and $\eta$ is a positive infinitesimal.
Therefore, the spectral function, which corresponds to the imaginary part of the one-body Green's function, can be obtained from
%
\begin{equation}
    A(\omega) = \sum_{i} \sum_{\lambda}A^i_{\lambda}A^{*i}_{\lambda}\delta(\omega-E_{\lambda})
\end{equation}
where $i$ is restricted to the 1-particle channel.
\section{Implementation}
\label{sec:implementation}
The effective Hamiltonian in Eq.~\ref{Heff:eq} is expressed in terms of the spin-orbitals.
Therefore, for closed-shell systems, the three-particle channel contains contributions from both spin doublets and spin quartets.
However, only the doublet contributions couple to the one-particle channel since the latter only contains spin doublets.
As a consequence, we transformed the Hamiltonian such that doublet and quartet states are separated.
The details of the spin adaptation of the Hamiltonian are given in Appendix \ref{app:spinadapt}.
The size $N^{SA}$ of the effective Hamiltonian for doublet states only is given by
\begin{equation}
\label{Eqn:NSA}
N^{SA}=2\left[n_V\frac{n_O (n_O+1)}{2}+n_O\frac{n_V (n_V+1)}{2} \right] + n_{Bas},
\end{equation}
where $n_O$ and $n_V$ are the number of occupied and virtual spatial orbitals, respectively, and $n_{Bas} = n_O + n_V$.
In Eq.~\eqref{Eqn:NSA}, the expression in brackets on the right-hand side is the total number of unique 3-particle states.
It is multiplied by 2 because 2 distinct spin doublets exist for each three-particle state. Finally, $n_{Bas}$ is the total number of 1-particle states.

It is numerically too expensive to diagonalise the effective Hamiltonian.
Therefore we use the Haydock-Lanczos method to calculate the spectral function~\cite{Hay72}.
This method follows the standard Lanczos algorithm in which the effective Hamiltonian is transformed into a real symmetric tridiagonal matrix $T$
with diagonal elements $a_i$ and off-diagonal elements $b_i$.
However, instead of diagonalising the tridiagonal matrix, we obtain the spectral function directly from a continued fraction according to
\begin{equation}
\label{Eqn:sf}
A(z)=-\frac{1}{\pi} \text{Im} \left[z - a_0 - \frac{b^2_1}{z - a_1 - \frac{b^2_2}{z - a_2 - ...}}\right]^{-1},
\end{equation}
with $z=\omega + i \eta$.
It is important to note that, in order to select the one-particle channel, the initial Lanczos vector should have non-zero elements only in the one-particle space.

Since the spectral function is obtained using the Lanczos-Haydock method, the bottleneck of a (3,1)-MCDE calculation is the construction of $\Sigma^{3 \text p}$ in the effective Hamiltonian.
The calculation of the hole-hole-electron channel of $\Sigma^{3 \text p}$ scales as $n_O^4 n_V + n_V^3 n_O^2$ while the calculation of its electron-electron-hole channel scales as $n_V^4 n_O + n_O^3 n_V^2$.
Therefore, the overall scaling of a (3,1)-MCDE calculation is $N^5$ with $N$ the number of electrons in the atom or molecule.

\section{Computational Details}
\label{sec:compdetails}
\begin{figure}[t]
\centering
\includegraphics[width=\columnwidth]{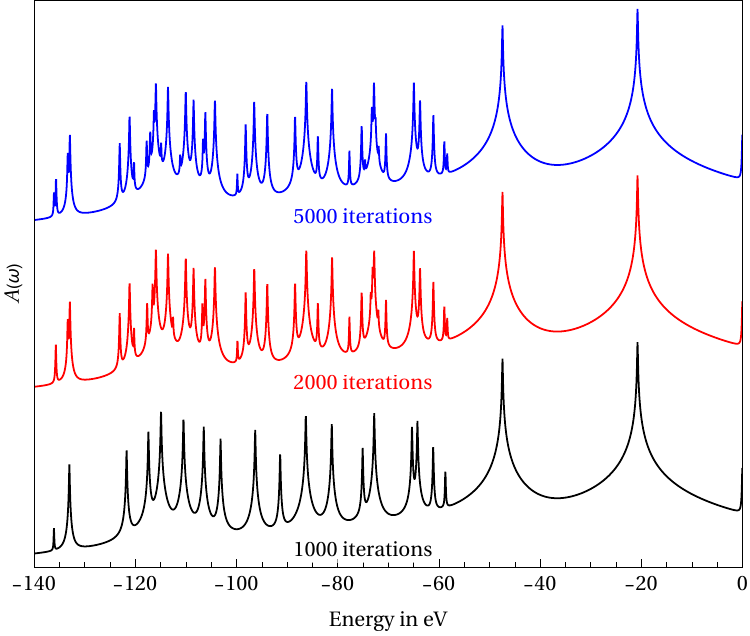}
\caption{The valence spectral functions of Neon obtained with the  aug-cc-pVQZ basis set using different numbers of Lanczos iterations.
The spectral functions are in logarithmic scale.}
\label{Fig:Lanczos}
\end{figure}

The geometries of the molecules were taken from Ref.~\cite{GW100}.
All Hartree-Fock calculations were performed with PySCF~\cite{Sun_2020}.
The restricted Hartree-Fock orbital energies as well as the two-electron integrals in terms of the Hartree-Fock orbitals were then read into our MCDE code to obtain the spectral functions.
The Lanczos iterations to obtain the (3,1)-MCDE spectral function were stopped when the norm of the Lanczos residual had become smaller than $10^{-12}$ Ha.
In practice this means a couple of thousand of iterations for each calculation.
The largest number of iterations to reach convergence was about 10 000 for the Neon atom using the aug-cc-pV6Z basis.
In Fig.~\ref{Fig:Lanczos} we show as an example the convergence of the valence spectral function of Neon using the aug-cc-pVQZ basis with the number of Lanczos iterations.
A lorentzian broadening was applied to all (3,1)-MCDE spectra by setting $\eta = 0.002$ Ha in Eq.~\eqref{Eqn:sf}.
We applied the same broadening to the FCI and $GW$ spectral functions.

The $GW$ spectral functions were obtained with MolGW~\cite{Bruneval_2016}.
We performed a perturbative $GW$ calculation on top of a restricted Hartree-Fock calculation.
The frequency-dependent $GW$ self-energy was calculated using the contour deformation approach.

The FCI calculations for the removal part of the spectral function were performed with PySCF. 
First, we calculated the FCI ground-state wave function $|\Psi_0\rangle$ and corresponding energy $E_0^N$ of the $N$-electron system.
Then, we calculate all ground and excited-state wave functions $|\Psi_i^{N-1}\rangle$ and corresponding energies $E_i^{N-1}$ of the system with $N-1$ electrons.
The removal energies and corresponding transition amplitudes are then calculated as $E_i^{N-1} - E_0^N$ and $\sum_j \left| \langle \Psi^{N-1}_i | a_j | \Psi_0^N \rangle\right|^2$, respectively.

In order to extrapolate the satellite energies to the complete basis set limit, one has to keep track of the excitation each satellite corresponds to in the various basis sets.
This information is stored in the eigenvectors of the effective Hamiltonian in Eq.~\eqref{Heff:eq} which contain coefficients corresponding to the one- and three-particle excitations.
We obtain approximate eigenvectors $v_i$ from the Lanczos vectors $q_i$ and the eigenvectors $t_i$ of the tridiagonal matrix $T$ according to
\begin{equation}
v_i = Q t_i
\end{equation}
where $Q$ is a matrix with $q_i$ as its columns.

In practice, we matched the biggest coefficients of the $v_i$ to identify equivalent satellites in two different basis sets.
Since the ordering of the RHF molecular orbitals (MO's) could change between two calculations with different basis sets, we calculated cross-overlap matrices between a given MO 
in the smaller basis set with all the MO's in the larger basis set.
We considered that the corresponding MO in the larger basis is the one that has the largest overlap with the MO of the smaller basis.
Finally, since the number of satellites becomes very large when increasing the size of the basis set, we only extrapolate energies corresponding to satellites with significant spectral weight.
\begin{figure*}[t]
\includegraphics[scale=.7]{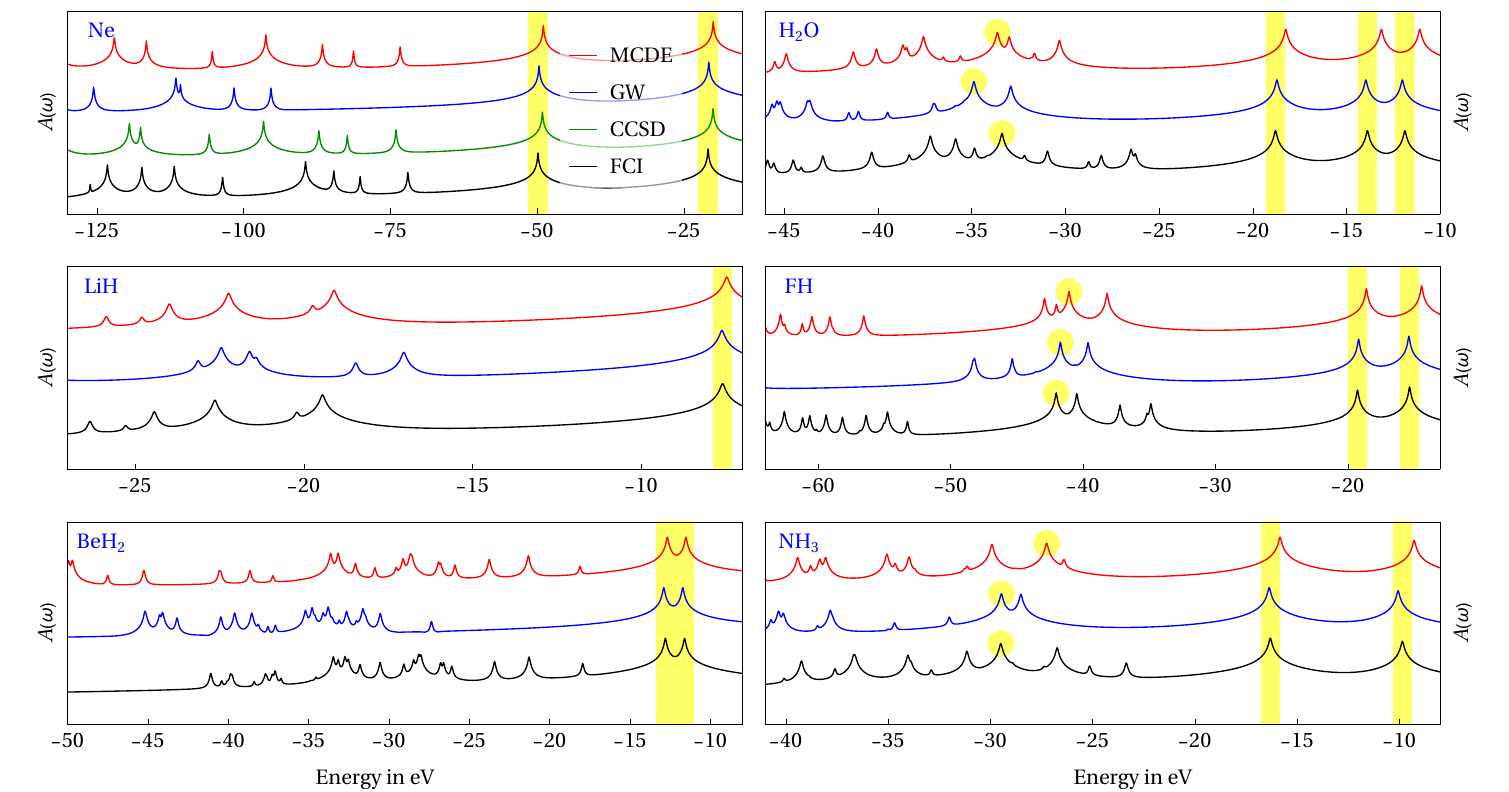}
\caption{The valence spectral functions of Neon and various small molecules calculated within FCI (black), GW (blue), and MCDE (red). 
For Neon, we also report the GFCCSD (green) spectral function taken from Ref.~\cite{Nishi_2018}.
The peaks highlighted in yellow correspond to the quasiparticle peaks.
Logarithmic scaling is used for the spectral function.
}
\label{fig:multiplot_fci}
\end{figure*}
\section{Results}
\label{sec:results}
In this section we will assess the accuracy and feasibility of (3,1)-MCDE calculations for atoms and molecules.
To judge the accuracy we will compare spectral functions obtained with the (3,1)-MCDE to those obtained within full configuration interaction (FCI) for a small basis set.
Then, to illustrate the feasibility of (3,1)-MCDE calculations, we will calculate the MCDE spectral functions for large basis sets.
Moreover, we will present a simple approach to obtain the MCDE spectral function in the complete basis-set limit.
\subsection{(3,1)-MCDE vs. FCI and $GW$}
For a given basis set, the FCI approach yields the exact ground and excited-state energies and wave functions.
From these solutions the exact spectral function corresponding to that basis set can be obtained.
Since FCI calculations scale exponentially with the number of basis functions, only calculations with few basis functions are numerically feasible.
Nevertheless, for small basis sets, FCI provides a useful reference against which the accuracy of the (3,1)-MCDE spectral functions can be benchmarked.

In Fig.~\ref{fig:multiplot_fci} we report the spectral functions of Neon and several small molecules obtained within the (3,1)-MCDE and compare them to the spectral functions obtained within FCI and $GW$.
In the case of Neon we also compare to a spectral function from Ref.~\cite{Nishi_2018} that was obtained 
using the coupled-cluster approach to the single-particle Green's function with single and double excitations (GFCCSD)~\cite{Nooijen_1992,Nooijen_1993,Nooijen_1995}.

For Neon we observe that the (3,1)-MCDE spectral function is in overall good agreement with the FCI spectrum and it is very similar to that obtained within GFCCSD.
The (3,1)-MCDE spectrum provides a remarkable improvement over the $GW$ spectral function which strongly overestimates the binding energy of the satellites.
For example, the visible satellite with the lowest binding energy in the FCI spectrum is at -51.2 eV while the (3,1)-MCDE predicts it to be at -53.3 eV.
Instead, $GW$ predicts this satellite to be at -74.6 eV, an error of more than 23 eV.

The spectral functions for the small molecules show a similar trend, i.e., $GW$ overestimates the binding energy of the satellites for all molecules except the smallest one, LiH, where it strongly underestimates them.
Instead, the (3,1)-MCDE yields spectral functions that are overall closer to those obtained within FCI, although also the (3,1)-MCDE overestimates the satellite binding energies except for LiH.
In particular, for LiH and BeH$_2$ the MCDE spectra compare very well to the FCI spectra.
Instead, the spectral functions obtained for HF, NH$_3$, and, especially, H$_2$O show larger differences with those obtained within FCI.
For those molecules, we also note that the (3,1)-MCDE underestimates the binding energies of most of the quasiparticles
Nevertheless, the (3,1)-MCDE spectral functions are overall in better agreement with the FCI spectra than those obtained with $GW$.

Obtaining FCI spectral functions for a wide frequency range is a numerical challenge using large basis sets.
Therefore, instead, in Table~\ref{Table:QP} we compare FCI valence and core quasiparticle energies from Refs.~\cite{Marie_2024} and \cite{Marie_2026_rel}
using the aug-cc-pVTZ (TZ) and aug-cc-pVQZ (QZ) basis sets to those obtained with the (3,1)-MCDE for various molecules.
The (3,1)-MCDE systematically underestimates the binding energies of the valence quasiparticles. This is consistent with the findings in Ref.~\cite{Keizer_2026}.
The underestimation ranges from 0.2 eV to 1 eV.
We observe the same underestimation of (3,1)-MCDE for the core quasiparticle energies, although care must be taken when comparing the core data since the FCI core quasiparticle energies were obtained using core-valence separation.
\begin{table}[h]
    \caption{Quasiparticle energies from the (3,1)-MCDE using the aug-cc-pVTZ (TZ) and aug-cc-pVQZ (QZ) basis sets for various molecules compared
    to valence FCI data from \cite{Marie_2024} and core FCI data from \cite{Marie_2026_rel}. All energies are in eV.}
\centering
\begin{tabular}{l|cc|cc}
\hline\hline
 & \multicolumn{2}{c|}{(3,1)-MCDE} & \multicolumn{2}{c}{FCI} \\
\text{Molecule} & \text{TZ} & \text{QZ} & \text{TZ} & \text{QZ} \\
\hline
\multirow{2}{*}{Ne Valence}
& {-20.605} & {-20.724} & {-21.461} & {-21.533} \\
& {-47.370} & {-47.413} & {-48.306} & {-48.340} \\
\hline
\multirow{3}{*}{H$_2$O Valence}
& {-11.897} & {-11.995} & {-12.679} & {-12.737} \\
& {-14.117} & {-14.204} & {-14.915} & {-14.962} \\
& {-18.384} & {-18.460} & {-19.024} & {-19.063} \\
\hline
\multirow{2}{*}{FH Valence}
& {-15.238} & {-15.336} & {-16.149} & {-16.214} \\
& {-19.287} & {-19.376} & {-20.054} & {-20.102} \\
\hline
\multirow{2}{*}{NH$_3$ Valence}
& {-10.308} & {-10.384} & {-10.901} & {-10.945} \\
& {-16.033} & {-16.098} & {-16.603} & {-16.640} \\
& -26.558 & -26.591 & -26.779 & -26.833  \\
\hline
\multirow{1}{*}{H$_2$O Core}
& {-538.12} & {-538.08} & {-539.87} & {-539.70} \\
\hline
\multirow{1}{*}{FH Core}
& {-692.15} & {-692.02} & {-694.10} & {-693.90} \\
\hline
\multirow{1}{*}{NH$_3$ Core}
& {-404.24} & {-404.30} & {-405.63} & {-405.49} \\
\hline\hline
\end{tabular}
\label{Table:QP}
\end{table}

The overestimation of the binding energy of the satellites could be due to the fact that not all the electron-hole and particle-particle interactions included in the three-particle self-energy $\Sigma^{3 \text p}$ are screened. 
Since screening effects are approximately proportional to the number of electrons, this omission could have important consequences for large molecules.
As we recently showed~\cite{Romaniello_2026}, all interactions can be properly screened by using the following three-particle self-energy
\begin{align}
    \Sigma^{3 \text p, scr}_{ijl;mok}&=  [(1 - f_i) (1 - f_j)f_l - f_if_j(1 - f_l)]
    \\ \nonumber &\times [\delta_{lk} (W_{ijom} - W_{ijmo})\nonumber + \delta_{mj}(W_{iklo} - v_{ikol}) 
    \\ \nonumber &+ \delta_{io} (W_{jklm} - v_{jkml}) - \delta_{oj} (W_{iklm}  - v_{ikml}) 
    \\ \nonumber &-  \delta_{im}  (W_{jklo} - v_{jkol})].
\end{align}
in which the matrix elements of the statically screened Coulomb potential $W(\vec{r}_1,\vec{r}_2)$  are given by
\begin{equation}\label{potential:eq}
    W_{ijlm}=\iint dx_1 dx_2 \varphi^*_i(x_1)\varphi^*_j(x_2)W(\vec{r}_1,\vec{r}_2)\varphi_l(x_2)\varphi_m(x_1).
\end{equation}
We have recently shown that the screened (3,1)-MCDE improves the quasiparticle energies of small molecules~\cite{Keizer_2026} showing its potential for improving also the spectral functions.
We defer the impact of the screened (3,1)-MCDE on the spectral functions of molecules to a future work.
Finally, we note that the screened potential has also been employed to accurately describe other types of many-body excitations~\cite{Deilmann_2016,Torche_2019,Sab22,Dis23-1,Venkatareddy_2025}.

So far, we have only used small basis sets because we wanted to compare to FCI spectra for which large basis sets are prohibitive.
In the next section we will calculate the MCDE spectral functions using larger basis sets and demonstrate how one can obtain photoemission spectra in the complete basis-set limit.
\subsection{Complete basis set limit}
\begin{figure}[t]
\centering
\includegraphics[width=\columnwidth]{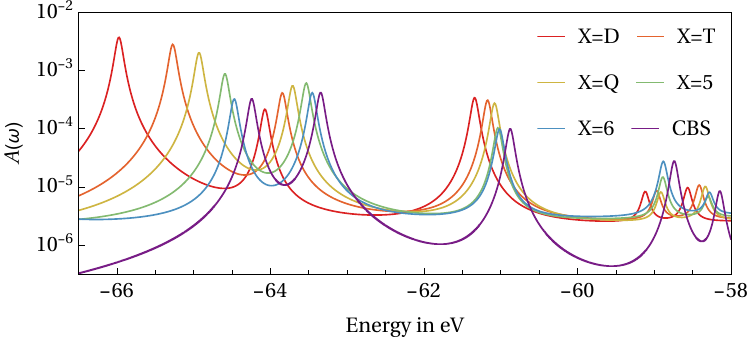}
\caption{The spectral function of the 5 neon satellite peaks with lowest binding energy calculated using various aug-cc-pVXZ basis sets. 
We note that the spectral function is in logarithmic scale.}
\label{fig:neon_extrapolation}
\end{figure}
\begin{figure}[b]
\includegraphics[width=\columnwidth]{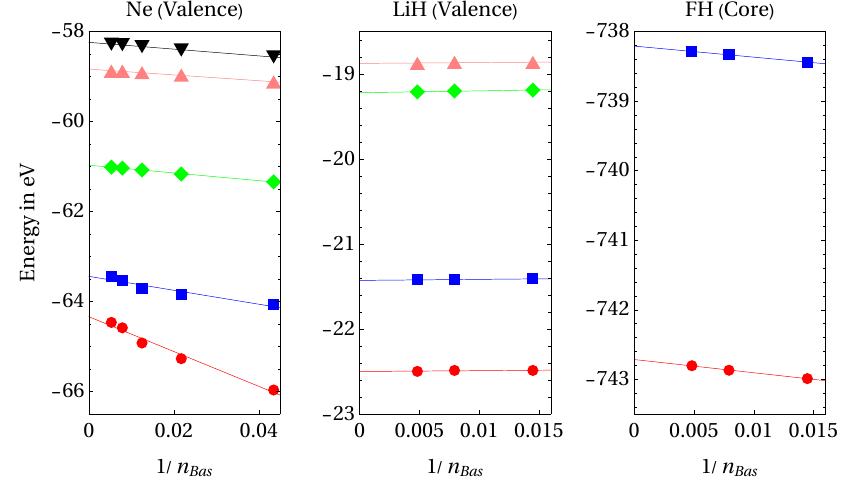}
\caption{
Valence satellite energies of Neon and LiH as well as core satellite energies of FH as a function of the inverse of the number of basis functions for various basis sets (from aug-cc-pVDZ to aug-cc-pV6Z for Neon and from aug-cc-pVTZ  to aug-cc-pV5Z  for FH and LiH).}
\label{fig:extrapolation_lines}
\end{figure}
\begin{figure*}[t]
\includegraphics[scale=.7]{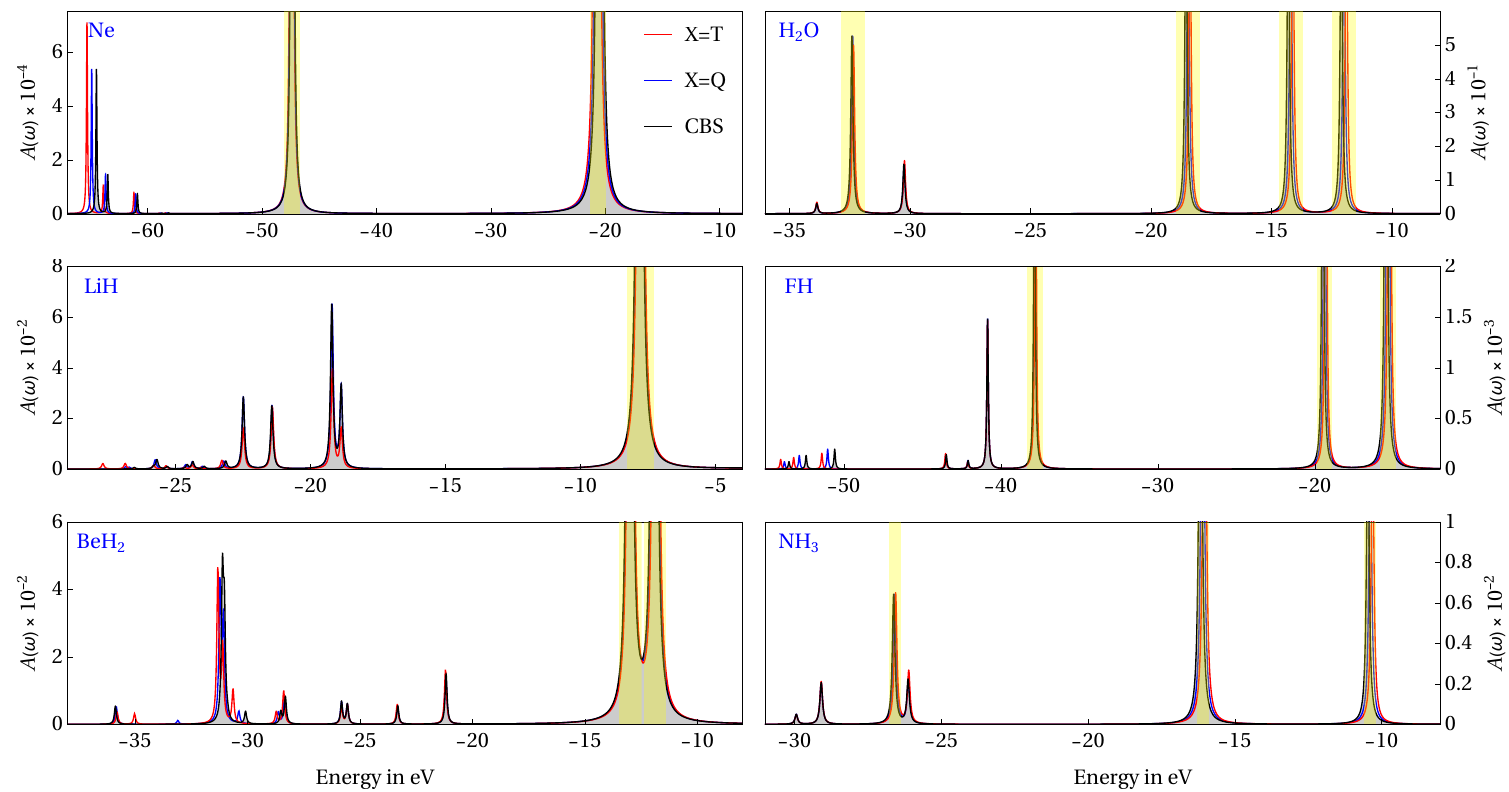}
\caption{The valence spectral functions of Neon and several molecules obtained by extrapolating to the complete basis-set limit using the aug-cc-pVTZ and aug-cc-pVQZ basis sets. The peaks highlighted in yellow correspond to the quasiparticle peaks.}
\label{fig:multiplot_ebl_qz}
\end{figure*}
The binding energies and intensities of the satellites converge extremely slowly with the size of the basis set.
As an example, in Fig. \ref{fig:neon_extrapolation} we report the spectral function of Neon for the aug-cc-pVXZ family of basis sets from double $\zeta$ (X=D) to sextuple $\zeta$ (X=6).
We only show the part of the spectrum containing the five satellite energies with the smallest binding energies.
We see that the closer a satellite is to its corresponding quasi-particle peak the faster is the convergence with basis-set size.
This was to be expected since low-lying satellites involve higher virtual orbitals.
However, even with the aug-cc-pV6Z basis set this part of the spectrum has not fully converged.
Since there are many more lower-lying satellites that are not included in Fig. \ref{fig:neon_extrapolation}, these observations lead to the conclusion that extremely large basis sets are required to converge the full spectral function.
However, calculations with such large basis sets are generally out of reach for most atoms and molecules.

Therefore, we propose a numerically much more efficient solution.
It has been shown that quasiparticle energies behave linearly as a function of the inverse of the number of basis functions and that, therefore, the complete basis set (CBS) limit for these energies can be obtained by extrapolation~\cite{vanSet13}.
Since both quasi-particle energies and satellite energies are electron removal and addition energies, a similar behaviour could exist for the satellite energies.
To verify this idea, we study the energies of the five Neon valence satellite energies shown in Fig.~\ref{fig:neon_extrapolation} as a function of the inverse of the number of basis functions.
We report the results in Fig.~\ref{fig:extrapolation_lines} and we clearly observe that the satellite energies indeed lie on straight lines.
We observe a similar behaviour for the satellite energies of the molecules.
In Fig.~\ref{fig:extrapolation_lines} we report the examples of the core satellite energies of FH and the valence satellite energies of LiH.
The root mean squared errors of the linear fits in Fig.~\ref{fig:extrapolation_lines} lie between 0.01 and 0.09 eV.
Therefore, as for the quasiparticle energies, we can also extrapolate the satellite energies to the CBS limit.
Unfortunately, the intensities of the satellite peaks do not scale linearly as a function of $1/n_{bas}$.
Therefore, we will use the intensities obtained within the largest basis set in the CBS spectral functions.
In Fig. \ref{fig:neon_extrapolation} we reported the spectral function of Neon containing five satellites in the CBS limit.
%

\begin{figure*}[t]
\includegraphics[scale=.7]{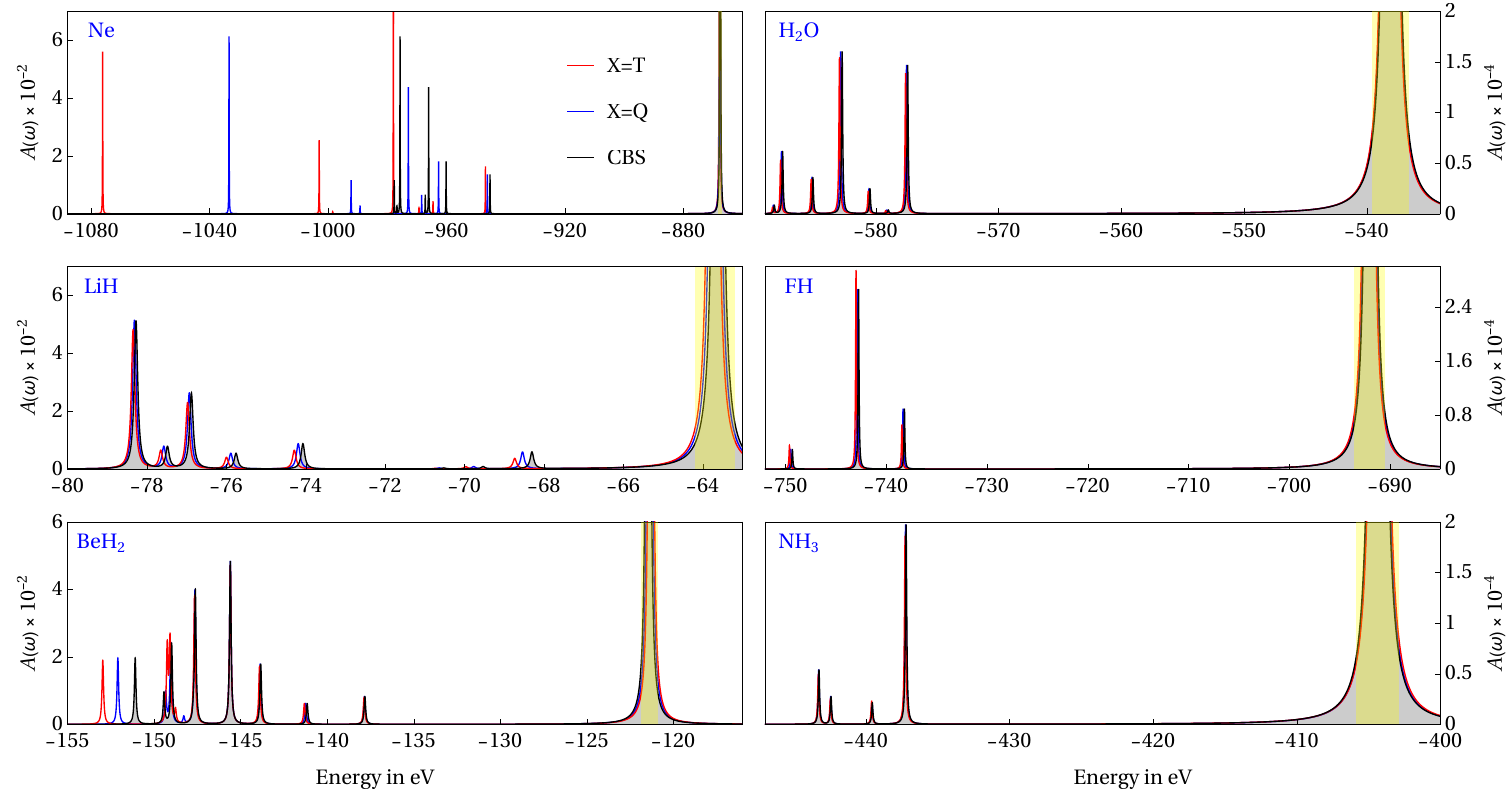}
\caption{The core spectral functions of Neon and several molecules obtained by extrapolating to the complete basis-set limit using the aug-cc-pVTZ and aug-cc-pVQZ basis sets.
The peaks highlighted in yellow correspond to the quasiparticle peaks.}
\label{fig:multiplot_core}
\end{figure*}

\begin{table}[t]
    \centering
    \caption{Five Neon satellite energies obtained from the extrapolation of the energies calculated using the aug-cc-pVXZ basis set for two consecutive values of X.}
    \begin{tabular}{c|ccccc}
		X& \multicolumn{5}{c}{\text{Neon satellite energies (eV)}}  \\
		\hline
		\hline
		\text{D+T} \, &  \, -64.55 \, & \, -63.59 \, & \, -60.97 \, & \, -58.79 \, & \, -58.24 \, \\
		\text{T+Q} & \, -64.36 \, & \, -63.45 \, & \, -60.87 \, & \, -58.75 \, & \, -58.15 \, \\
		\text{Q+5} & -63.85 & -63.05 & -60.80 & -58.66 & -58.06 \\
		\text{5+6} & -64.07 & -63.14 & -60.82 & -58.73 & -58.11 \\
		\hline
	\end{tabular}
	\label{tab:extrapolation_table}
\end{table}

In Table \ref{tab:extrapolation_table} we report five Neon satellite energies obtained from the extrapolation of two energies obtained with aug-cc-pVXZ basis sets using two consecutive values of X.
The extrapolations from different basis sets of the three Neon satellites with the lowest binding energy are converged to about 0.1 eV compared to the extrapolation from 5Z and 6Z basis sets.
In particular, the extrapolation from the TZ and QZ basis sets yields energies that are very close to those obtained from the 5Z/6Z extrapolation.
Moreover, the aug-cc-pVQZ basis set remains affordable for larger molecules.
For these reasons we will, in the following, use the extrapolation from the aug-cc-pVTZ and aug-cc-pVQZ basis sets.
Finally, we note that, in general, the accuracy of the extrapolation decreases when the binding energy of the satellite increases.
In Fig.~\ref{fig:multiplot_ebl_qz} we report the (3,1)-MCDE valence spectral functions thus obtained for Ne, LiH, BeH$_2$, H$_2$O, FH and NH$_3$.
We observe that, for the given energy ranges, the Neon and FH spectral functions are not converged using the aug-cc-pVQZ basis and an extrapolation is crucial.

Finally, in Fig.~\ref{fig:multiplot_core} we report the (3,1)-MCDE core spectral functions for Ne, LiH, BeH$_2$, H$_2$O, FH and NH$_3$ obtained using the same extrapolation method.
Again we observe that, for the given energy ranges, the spectral functions of some systems and, in particular Neon, are not converged using the aug-cc-pVQZ basis and an extrapolation is needed.
\section{Conclusions}
\label{sec:conclusions}
We have implemented the (3,1)-MCDE for finite systems and used it to calculate the photoemission spectral functions of atoms and molecules.
For small basis sets the (3,1)-MCDE spectra are in better agreement with FCI spectra than those obtained with $GW$.
However, for some molecules significant differences remain for the binding energies of the satellites.
This is could be due to a lack of screening of the electron-hole and particle-particle excitations in the (3,1)-MCDE.
In future work we will investigate the performance of the screened (3,1)-MCDE, in which these excitations are properly screened, for the spectral functions of atoms and molecules.

We have also demonstrated that the binding energies of the satellites converge extremely slowly with the size of the basis set.
We have presented a simple extrapolation technique to obtain spectral functions in the complete basis-set limit.

Finally, our implementation of the (3,1)-MCDE for atoms and molecules sets the stage for the implementation of other multichannel Dyson equations that describe other types of spectrosopies~\cite{riva_prb_25,Sellie_2026}.
\section*{Acknowledgment}
We thank the French Agence Nationale de la Recherche (ANR) for financial support (Grant Agreement ANR-22-CE29-0001 and ANR-22-CE30-0027).
We thank Fabien Bruneval, Arno F\"orster, Mike Keizer and Carlo Barbieri for fruitful discussions.
\appendix
\section{Spin-adapted (3,1)-MCDE equations}
As mentioned in the main text, the three-particle channel of the effective Hamiltonian in Eq.~\ref{Heff:eq} contains contributions from both spin doublets and spin quartets.
However, only the doublet contributions couple to the one-particle channel since the latter only contains spin doublets.
In order to separate the doublet and quartet states we can perform a basis transformation.
For the spin up channel this transformation is given by~\cite{Nie84}
\begin{align}
A_i^{D} & = A_{i\uparrow} = \epsilon_i
\\
A_{ijl}^{D1} &= \frac{1}{\sqrt{2}} (A_{i\uparrow j\downarrow l\downarrow} - A_{i\downarrow j\uparrow l\downarrow})
\\
A_{ijl}^{D2} &= \frac{1}{\sqrt{6}} (2 A_{i\uparrow j\uparrow l\uparrow} + A_{i\uparrow j\downarrow l\downarrow} + A_{i\downarrow j\uparrow l\downarrow})
\\
A_{ijl}^{Q} &= \frac{1}{\sqrt{3}} (- A_{i\uparrow j\uparrow l\uparrow} + A_{i\uparrow j\downarrow l\downarrow} + A_{i\downarrow j\uparrow l\downarrow})
\\
\end{align}
Within this basis, the spin-up channel of the effective Hamiltonian for the doublet states $H_{3}^{D,\uparrow}$ becomes
\begin{equation}
	\label{eq:equation47}
	H_{3}^{D,\uparrow} = 
	\begin{pmatrix}
		\epsilon^{\text{1p}} &C^{D1} & C^{D2}  \\
		\left[C^{D1}\right]^{\dagger} & \epsilon^{\text{3p}} + H^{D1}  & C^{D1,D2}\\
		\left[C^{D2}\right]^{\dagger}&  \left[C^{D1,D2}\right]^{\dagger} &  \epsilon^{\text{3p}} + H^{D2} \\
	\end{pmatrix}
\end{equation}
in which
\label{app:spinadapt}
\begin{widetext}
\begin{align}
	\epsilon^{\text{1p}}_{i,m} &= \delta_{im} \epsilon_i
	\\
         \epsilon^{\text{3p}}_{i\ge jl;m\ge ok} &= \delta_{im}\delta_{jo}\delta_{lk} (\epsilon_i + \epsilon_j - \epsilon_l)
         \\
	C^{D1}_{i;m\ge ok} &= \left[\sqrt{\frac12}\right]^{\delta_{mo}} \sqrt{\frac12} (f_{m}-f_{k}) (f_{o}-f_{k}) [\mathcal{V}_{ikom} + \mathcal{V}_{ikmo}]
	\\
	C^{D2}_{i;m\ge ok} &= \sqrt{\frac32} (f_{m}-f_{k}) (f_{o}-f_{k}) [\mathcal{V}_{ikom} - \mathcal{V}_{ikmo}]
	\\
	\left[C^{D1}\right]^{\dagger}_{i\ge jl;m} &= \left[\sqrt{\frac12}\right]^{\delta_{ij}}\sqrt{\frac12} (f_{i}-f_{l})(f_{j}-f_{l}) [\mathcal{V}_{ijlm} + \mathcal{V}_{ijml}]
	\\
	\left[C^{D2}\right]^{\dagger}_{i\ge jl;m}  &= \sqrt{\frac32} (f_{i}-f_{l})(f_{j}-f_{l}) [\mathcal{V}_{ijlm} - \mathcal{V}_{ijml}]
	\\
	H^{D1}_{i\ge jl;m\ge ok} & = 
	\left[\sqrt{\frac12}\right]^{\delta_{mo}} 
	\left[\sqrt{\frac12}\right]^{\delta_{ij}} 
	\left[(1\!-\!f_i)\!(1\!-\!f_j)f_l\!-\!f_if_j(1\!-\!f_l)\right]
	\bigg(\delta_{lk} (\mathcal{V}_{ijom} + \mathcal{V}_{ijmo}) -
	\delta_{mj} (\mathcal{V}_{iklo} - \frac12 \mathcal{V}_{ikol}) 
	\nonumber \\ & - \delta_{io} (\mathcal{V}_{jklm} - \frac12 \mathcal{V}_{jkml}) 
	- \delta_{oj} (\mathcal{V}_{iklm} - \frac12 \mathcal{V}_{ikml}) - \delta_{im} (\mathcal{V}_{jklo} - \frac12 \mathcal{V}_{jkol}) \bigg)
	\\
	H^{D2}_{i\ge jl;m\ge ok}  & = \left[(1\!-\!f_i)\!(1\!-\!f_j)f_l\!-\!f_if_j(1\!-\!f_l)\right]\bigg(\delta_{lk} (\mathcal{V}_{ijom} - \mathcal{V}_{ijmo}) +
	\delta_{mj} (\mathcal{V}_{iklo} 
	\nonumber \\ & - \frac32 \mathcal{V}_{ikol}) + \delta_{io} (\mathcal{V}_{jklm} - \frac32 \mathcal{V}_{jkml}) 
	- \delta_{oj} (\mathcal{V}_{iklm} - \frac32 \mathcal{V}_{ikml}) - \delta_{im} (\mathcal{V}_{jklo} - \frac32 \mathcal{V}_{jkol}) \bigg)
	\\ 
	C^{D1,D2}_{i\ge jl;m\ge ok}  & = \left[\sqrt{\frac12}\right]^{\delta_{ij}} \frac{\sqrt{3}}{2} \left[(1\!-\!f_i)\!(1\!-\!f_j)f_l\!-\!f_if_j(1\!-\!f_l)\right]
	[\delta_{mj} \mathcal{V}_{ikol} - \delta_{io}  \mathcal{V}_{jkml} - \delta_{oj} \mathcal{V}_{ikml} + \delta_{im} \mathcal{V}_{jkol}]
	\\ 
	\left[C^{D1,D2}\right]^{\dagger}_{i\ge jl;m\ge ok} & = \left[\sqrt{\frac12}\right]^{\delta_{mo}} \frac{\sqrt{3}}{2} \left[(1\!-\!f_i)\!(1\!-\!f_j)f_l\!-\!f_if_j(1\!-\!f_l)\right]
	 [-\delta_{mj} \mathcal{V}_{ikol} + \delta_{io}  \mathcal{V}_{jkml} - \delta_{oj} \mathcal{V}_{ikml} + \delta_{im} \mathcal{V}_{jkol}].
\end{align}
\end{widetext}
Here all indices refer to spatial orbitals and where
\begin{equation}
\mathcal{V}_{ijlm}=\iint d\vec{r_1} d\vec{r_2}  \phi^\ast_i(\vec{r}_1)\phi^\ast_j(\vec{r}_2)v(\vec{r}_1, \vec{r}_2)\phi_l(\vec{r}_2)\phi_m(\vec{r}_1),
\end{equation}
in which $\phi_i(\vec{r})$ is a Hartree-Fock spatial orbital.
A similar transformation can be done for the spin-down channel.


\begin{thebibliography}{69}%
\makeatletter
\providecommand \@ifxundefined [1]{%
 \@ifx{#1\undefined}
}%
\providecommand \@ifnum [1]{%
 \ifnum #1\expandafter \@firstoftwo
 \else \expandafter \@secondoftwo
 \fi
}%
\providecommand \@ifx [1]{%
 \ifx #1\expandafter \@firstoftwo
 \else \expandafter \@secondoftwo
 \fi
}%
\providecommand \natexlab [1]{#1}%
\providecommand \enquote  [1]{``#1''}%
\providecommand \bibnamefont  [1]{#1}%
\providecommand \bibfnamefont [1]{#1}%
\providecommand \citenamefont [1]{#1}%
\providecommand \href@noop [0]{\@secondoftwo}%
\providecommand \href [0]{\begingroup \@sanitize@url \@href}%
\providecommand \@href[1]{\@@startlink{#1}\@@href}%
\providecommand \@@href[1]{\endgroup#1\@@endlink}%
\providecommand \@sanitize@url [0]{\catcode `\\12\catcode `\$12\catcode
  `\&12\catcode `\#12\catcode `\^12\catcode `\_12\catcode `\%12\relax}%
\providecommand \@@startlink[1]{}%
\providecommand \@@endlink[0]{}%
\providecommand \url  [0]{\begingroup\@sanitize@url \@url }%
\providecommand \@url [1]{\endgroup\@href {#1}{\urlprefix }}%
\providecommand \urlprefix  [0]{URL }%
\providecommand \Eprint [0]{\href }%
\providecommand \doibase [0]{https://doi.org/}%
\providecommand \selectlanguage [0]{\@gobble}%
\providecommand \bibinfo  [0]{\@secondoftwo}%
\providecommand \bibfield  [0]{\@secondoftwo}%
\providecommand \translation [1]{[#1]}%
\providecommand \BibitemOpen [0]{}%
\providecommand \bibitemStop [0]{}%
\providecommand \bibitemNoStop [0]{.\EOS\space}%
\providecommand \EOS [0]{\spacefactor3000\relax}%
\providecommand \BibitemShut  [1]{\csname bibitem#1\endcsname}%
\let\auto@bib@innerbib\@empty
\bibitem [{\citenamefont {Hüfner}(2003)}]{hufner}%
  \BibitemOpen
  \bibfield  {author} {\bibinfo {author} {\bibfnamefont {S.}~\bibnamefont
  {Hüfner}},\ }\href@noop {} {\emph {\bibinfo {title} {Photoelectron
  Spectroscopy: principles and applications}}}\ (\bibinfo  {publisher}
  {Springer},\ \bibinfo {year} {2003})\BibitemShut {NoStop}%
\bibitem [{\citenamefont {Fischer}\ and\ \citenamefont
  {Pratt}(2022)}]{Fischer_2022}%
  \BibitemOpen
  \bibfield  {author} {\bibinfo {author} {\bibfnamefont {I.}~\bibnamefont
  {Fischer}}\ and\ \bibinfo {author} {\bibfnamefont {S.~T.}\ \bibnamefont
  {Pratt}},\ }\bibfield  {title} {\bibinfo {title} {Photoelectron spectroscopy
  in molecular physical chemistry},\ }\href
  {https://doi.org/10.1039/d1cp04984d} {\bibfield  {journal} {\bibinfo
  {journal} {Physical Chemistry Chemical Physics}\ }\textbf {\bibinfo {volume}
  {24}},\ \bibinfo {pages} {1944} (\bibinfo {year} {2022})}\BibitemShut
  {NoStop}%
\bibitem [{\citenamefont {de~Groot}(2001)}]{deGroot_2001}%
  \BibitemOpen
  \bibfield  {author} {\bibinfo {author} {\bibfnamefont {F.}~\bibnamefont
  {de~Groot}},\ }\bibfield  {title} {\bibinfo {title} {High-resolution x-ray
  emission and x-ray absorption spectroscopy},\ }\href
  {https://doi.org/10.1021/cr9900681} {\bibfield  {journal} {\bibinfo
  {journal} {Chemical Reviews}\ }\textbf {\bibinfo {volume} {101}},\ \bibinfo
  {pages} {1779} (\bibinfo {year} {2001})}\BibitemShut {NoStop}%
\bibitem [{\citenamefont {de~Groot}\ and\ \citenamefont
  {Kotani}(2008)}]{deGroot_2008}%
  \BibitemOpen
  \bibfield  {author} {\bibinfo {author} {\bibfnamefont {F.}~\bibnamefont
  {de~Groot}}\ and\ \bibinfo {author} {\bibfnamefont {A.}~\bibnamefont
  {Kotani}},\ }\href@noop {} {\emph {\bibinfo {title} {Core Level Spectroscopy
  of Solids}}},\ Advances in Condensed Matter Science\ (\bibinfo  {publisher}
  {CRC Press},\ \bibinfo {address} {Boca Raton},\ \bibinfo {year}
  {2008})\BibitemShut {NoStop}%
\bibitem [{\citenamefont {van~der Heide}(2011)}]{vanderHeide_2011}%
  \BibitemOpen
  \bibfield  {author} {\bibinfo {author} {\bibfnamefont {P.}~\bibnamefont
  {van~der Heide}},\ }\href@noop {} {\emph {\bibinfo {title} {Photoelectron
  Spectroscopy: An Introduction to Principles and Practices}}},\ Wiley Series
  on Mass Spectrometry\ (\bibinfo  {publisher} {John Wiley and Sons},\ \bibinfo
  {address} {Chichester, West Sussex, Hoboken, NJ},\ \bibinfo {year}
  {2011})\BibitemShut {NoStop}%
\bibitem [{\citenamefont {Stevie}\ and\ \citenamefont
  {Donley}(2020)}]{Stevie_2020}%
  \BibitemOpen
  \bibfield  {author} {\bibinfo {author} {\bibfnamefont {F.~A.}\ \bibnamefont
  {Stevie}}\ and\ \bibinfo {author} {\bibfnamefont {C.~L.}\ \bibnamefont
  {Donley}},\ }\bibfield  {title} {\bibinfo {title} {Introduction to x-ray
  photoelectron spectroscopy},\ }\href {https://doi.org/10.1116/6.0000412}
  {\bibfield  {journal} {\bibinfo  {journal} {Journal of Vacuum Science and
  Technology A}\ }\textbf {\bibinfo {volume} {38}},\ \bibinfo {pages} {063204}
  (\bibinfo {year} {2020})}\BibitemShut {NoStop}%
\bibitem [{\citenamefont {Bagus}\ \emph {et~al.}(2018)\citenamefont {Bagus},
  \citenamefont {Ilton},\ and\ \citenamefont {Nelin}}]{Bagus_2018}%
  \BibitemOpen
  \bibfield  {author} {\bibinfo {author} {\bibfnamefont {P.~S.}\ \bibnamefont
  {Bagus}}, \bibinfo {author} {\bibfnamefont {E.}~\bibnamefont {Ilton}},\ and\
  \bibinfo {author} {\bibfnamefont {C.~J.}\ \bibnamefont {Nelin}},\ }\bibfield
  {title} {\bibinfo {title} {Extracting chemical information from xps spectra:
  a perspective},\ }\href {https://doi.org/10.1007/s10562-018-2417-1}
  {\bibfield  {journal} {\bibinfo  {journal} {Catalysis Letters}\ }\textbf
  {\bibinfo {volume} {148}},\ \bibinfo {pages} {1785} (\bibinfo {year}
  {2018})}\BibitemShut {NoStop}%
\bibitem [{\citenamefont {Hedin}(1965)}]{Hed65}%
  \BibitemOpen
  \bibfield  {author} {\bibinfo {author} {\bibfnamefont {L.}~\bibnamefont
  {Hedin}},\ }\bibfield  {title} {\bibinfo {title} {New method for calculating
  the one-particle green's function with application to the electron-gas
  problem},\ }\href {https://doi.org/10.1103/PhysRev.139.A796} {\bibfield
  {journal} {\bibinfo  {journal} {Phys. Rev.}\ }\textbf {\bibinfo {volume}
  {139}},\ \bibinfo {pages} {A796} (\bibinfo {year} {1965})}\BibitemShut
  {NoStop}%
\bibitem [{\citenamefont {Hybertsen}\ and\ \citenamefont
  {Louie}(1986)}]{Hybertsen_1986}%
  \BibitemOpen
  \bibfield  {author} {\bibinfo {author} {\bibfnamefont {M.~S.}\ \bibnamefont
  {Hybertsen}}\ and\ \bibinfo {author} {\bibfnamefont {S.~G.}\ \bibnamefont
  {Louie}},\ }\bibfield  {title} {\bibinfo {title} {Electron correlation in
  semiconductors and insulators: Band gaps and quasiparticle energies},\ }\href
  {https://doi.org/10.1103/PhysRevB.34.5390} {\bibfield  {journal} {\bibinfo
  {journal} {Phys. Rev. B}\ }\textbf {\bibinfo {volume} {34}},\ \bibinfo
  {pages} {5390} (\bibinfo {year} {1986})}\BibitemShut {NoStop}%
\bibitem [{\citenamefont {Aryasetiawan}\ and\ \citenamefont
  {Gunnarsson}(1998)}]{Ary98}%
  \BibitemOpen
  \bibfield  {author} {\bibinfo {author} {\bibfnamefont {F.}~\bibnamefont
  {Aryasetiawan}}\ and\ \bibinfo {author} {\bibfnamefont {O.}~\bibnamefont
  {Gunnarsson}},\ }\bibfield  {title} {\bibinfo {title} {The gw method},\
  }\href {https://doi.org/10.1088/0034-4885/61/3/002} {\bibfield  {journal}
  {\bibinfo  {journal} {Reports on Progress in Physics}\ }\textbf {\bibinfo
  {volume} {61}},\ \bibinfo {pages} {237} (\bibinfo {year} {1998})}\BibitemShut
  {NoStop}%
\bibitem [{\citenamefont {Hedin}(1999)}]{Hed99}%
  \BibitemOpen
  \bibfield  {author} {\bibinfo {author} {\bibfnamefont {L.}~\bibnamefont
  {Hedin}},\ }\bibfield  {title} {\bibinfo {title} {On correlation effects in
  electron spectroscopies and the gw approximation},\ }\href
  {https://doi.org/10.1088/0953-8984/11/42/201} {\bibfield  {journal} {\bibinfo
   {journal} {Journal of Physics: Condensed Matter}\ }\textbf {\bibinfo
  {volume} {11}},\ \bibinfo {pages} {R489} (\bibinfo {year}
  {1999})}\BibitemShut {NoStop}%
\bibitem [{\citenamefont {Reining}(2018)}]{Reining_2018}%
  \BibitemOpen
  \bibfield  {author} {\bibinfo {author} {\bibfnamefont {L.}~\bibnamefont
  {Reining}},\ }\bibfield  {title} {\bibinfo {title} {The {$GW$} approximation:
  content, successes and limitations},\ }\href
  {https://doi.org/https://doi.org/10.1002/wcms.1344} {\bibfield  {journal}
  {\bibinfo  {journal} {WIREs Computational Molecular Science}\ }\textbf
  {\bibinfo {volume} {8}},\ \bibinfo {pages} {e1344} (\bibinfo {year}
  {2018})}\BibitemShut {NoStop}%
\bibitem [{\citenamefont {Golze}\ \emph {et~al.}(2019)\citenamefont {Golze},
  \citenamefont {Dvorak},\ and\ \citenamefont {Rinke}}]{Golze_2019}%
  \BibitemOpen
  \bibfield  {author} {\bibinfo {author} {\bibfnamefont {D.}~\bibnamefont
  {Golze}}, \bibinfo {author} {\bibfnamefont {M.}~\bibnamefont {Dvorak}},\ and\
  \bibinfo {author} {\bibfnamefont {P.}~\bibnamefont {Rinke}},\ }\bibfield
  {title} {\bibinfo {title} {The {$GW$} compendium: A practical guide to
  theoretical photoemission spectroscopy},\ }\bibfield  {journal} {\bibinfo
  {journal} {Frontiers in Chemistry}\ }\href
  {https://doi.org/10.3389/fchem.2019.00377} {10.3389/fchem.2019.00377}
  (\bibinfo {year} {2019})\BibitemShut {NoStop}%
\bibitem [{\citenamefont {van Schilfgaarde}\ \emph {et~al.}(2006)\citenamefont
  {van Schilfgaarde}, \citenamefont {Kotani},\ and\ \citenamefont
  {Faleev}}]{VanS06}%
  \BibitemOpen
  \bibfield  {author} {\bibinfo {author} {\bibfnamefont {M.}~\bibnamefont {van
  Schilfgaarde}}, \bibinfo {author} {\bibfnamefont {T.}~\bibnamefont
  {Kotani}},\ and\ \bibinfo {author} {\bibfnamefont {S.}~\bibnamefont
  {Faleev}},\ }\bibfield  {title} {\bibinfo {title} {Quasiparticle
  self-consistent $gw$ theory},\ }\href
  {https://doi.org/10.1103/PhysRevLett.96.226402} {\bibfield  {journal}
  {\bibinfo  {journal} {Phys. Rev. Lett.}\ }\textbf {\bibinfo {volume} {96}},\
  \bibinfo {pages} {226402} (\bibinfo {year} {2006})}\BibitemShut {NoStop}%
\bibitem [{\citenamefont {van Setten}\ \emph {et~al.}(2015)\citenamefont {van
  Setten}, \citenamefont {Caruso}, \citenamefont {Sharifzadeh}, \citenamefont
  {Ren}, \citenamefont {Scheffler}, \citenamefont {Liu}, \citenamefont
  {Lischner}, \citenamefont {Lin}, \citenamefont {Deslippe}, \citenamefont
  {Louie}, \citenamefont {Yang}, \citenamefont {Weigend}, \citenamefont
  {Neaton}, \citenamefont {Evers},\ and\ \citenamefont {Rinke}}]{GW100}%
  \BibitemOpen
  \bibfield  {author} {\bibinfo {author} {\bibfnamefont {M.~J.}\ \bibnamefont
  {van Setten}}, \bibinfo {author} {\bibfnamefont {F.}~\bibnamefont {Caruso}},
  \bibinfo {author} {\bibfnamefont {S.}~\bibnamefont {Sharifzadeh}}, \bibinfo
  {author} {\bibfnamefont {X.}~\bibnamefont {Ren}}, \bibinfo {author}
  {\bibfnamefont {M.}~\bibnamefont {Scheffler}}, \bibinfo {author}
  {\bibfnamefont {F.}~\bibnamefont {Liu}}, \bibinfo {author} {\bibfnamefont
  {J.}~\bibnamefont {Lischner}}, \bibinfo {author} {\bibfnamefont
  {L.}~\bibnamefont {Lin}}, \bibinfo {author} {\bibfnamefont {J.~R.}\
  \bibnamefont {Deslippe}}, \bibinfo {author} {\bibfnamefont {S.~G.}\
  \bibnamefont {Louie}}, \bibinfo {author} {\bibfnamefont {C.}~\bibnamefont
  {Yang}}, \bibinfo {author} {\bibfnamefont {F.}~\bibnamefont {Weigend}},
  \bibinfo {author} {\bibfnamefont {J.~B.}\ \bibnamefont {Neaton}}, \bibinfo
  {author} {\bibfnamefont {F.}~\bibnamefont {Evers}},\ and\ \bibinfo {author}
  {\bibfnamefont {P.}~\bibnamefont {Rinke}},\ }\bibfield  {title} {\bibinfo
  {title} {Gw100: Benchmarking g0w0 for molecular systems},\ }\href
  {https://doi.org/10.1021/acs.jctc.5b00453} {\bibfield  {journal} {\bibinfo
  {journal} {Journal of Chemical Theory and Computation}\ }\textbf {\bibinfo
  {volume} {11}},\ \bibinfo {pages} {5665} (\bibinfo {year}
  {2015})}\BibitemShut {NoStop}%
\bibitem [{\citenamefont {Li}\ \emph {et~al.}(2022)\citenamefont {Li},
  \citenamefont {Jin}, \citenamefont {Rinke}, \citenamefont {Yang},\ and\
  \citenamefont {Golze}}]{Li_2022}%
  \BibitemOpen
  \bibfield  {author} {\bibinfo {author} {\bibfnamefont {J.}~\bibnamefont
  {Li}}, \bibinfo {author} {\bibfnamefont {Y.}~\bibnamefont {Jin}}, \bibinfo
  {author} {\bibfnamefont {P.}~\bibnamefont {Rinke}}, \bibinfo {author}
  {\bibfnamefont {W.}~\bibnamefont {Yang}},\ and\ \bibinfo {author}
  {\bibfnamefont {D.}~\bibnamefont {Golze}},\ }\bibfield  {title} {\bibinfo
  {title} {Benchmark of gw methods for core-level binding energies},\ }\href
  {https://doi.org/10.1021/acs.jctc.2c00617} {\bibfield  {journal} {\bibinfo
  {journal} {Journal of Chemical Theory and Computation}\ }\textbf {\bibinfo
  {volume} {18}},\ \bibinfo {pages} {7570} (\bibinfo {year}
  {2022})}\BibitemShut {NoStop}%
\bibitem [{\citenamefont {Langreth}(1970)}]{Lan70}%
  \BibitemOpen
  \bibfield  {author} {\bibinfo {author} {\bibfnamefont {D.~C.}\ \bibnamefont
  {Langreth}},\ }\bibfield  {title} {\bibinfo {title} {Singularities in the
  x-ray spectra of metals},\ }\href@noop {} {\bibfield  {journal} {\bibinfo
  {journal} {Physical Review B}\ }\textbf {\bibinfo {volume} {1}},\ \bibinfo
  {pages} {471} (\bibinfo {year} {1970})}\BibitemShut {NoStop}%
\bibitem [{\citenamefont {Aryasetiawan}\ \emph {et~al.}(1996)\citenamefont
  {Aryasetiawan}, \citenamefont {Hedin},\ and\ \citenamefont
  {Karlsson}}]{Ary96}%
  \BibitemOpen
  \bibfield  {author} {\bibinfo {author} {\bibfnamefont {F.}~\bibnamefont
  {Aryasetiawan}}, \bibinfo {author} {\bibfnamefont {L.}~\bibnamefont
  {Hedin}},\ and\ \bibinfo {author} {\bibfnamefont {K.}~\bibnamefont
  {Karlsson}},\ }\bibfield  {title} {\bibinfo {title} {Multiple plasmon
  satellites in na and al spectral functions from ab initio cumulant
  expansion},\ }\href {https://doi.org/10.1103/PhysRevLett.77.2268} {\bibfield
  {journal} {\bibinfo  {journal} {Phys. Rev. Lett.}\ }\textbf {\bibinfo
  {volume} {77}},\ \bibinfo {pages} {2268} (\bibinfo {year}
  {1996})}\BibitemShut {NoStop}%
\bibitem [{\citenamefont {Guzzo}\ \emph {et~al.}(2011)\citenamefont {Guzzo},
  \citenamefont {Lani}, \citenamefont {Sottile}, \citenamefont {Romaniello},
  \citenamefont {Gatti}, \citenamefont {Kas}, \citenamefont {Rehr},
  \citenamefont {Silly}, \citenamefont {Sirotti},\ and\ \citenamefont
  {Reining}}]{Guz11}%
  \BibitemOpen
  \bibfield  {author} {\bibinfo {author} {\bibfnamefont {M.}~\bibnamefont
  {Guzzo}}, \bibinfo {author} {\bibfnamefont {G.}~\bibnamefont {Lani}},
  \bibinfo {author} {\bibfnamefont {F.}~\bibnamefont {Sottile}}, \bibinfo
  {author} {\bibfnamefont {P.}~\bibnamefont {Romaniello}}, \bibinfo {author}
  {\bibfnamefont {M.}~\bibnamefont {Gatti}}, \bibinfo {author} {\bibfnamefont
  {J.~J.}\ \bibnamefont {Kas}}, \bibinfo {author} {\bibfnamefont {J.~J.}\
  \bibnamefont {Rehr}}, \bibinfo {author} {\bibfnamefont {M.~G.}\ \bibnamefont
  {Silly}}, \bibinfo {author} {\bibfnamefont {F.}~\bibnamefont {Sirotti}},\
  and\ \bibinfo {author} {\bibfnamefont {L.}~\bibnamefont {Reining}},\
  }\bibfield  {title} {\bibinfo {title} {Valence electron photoemission
  spectrum of semiconductors: Ab initio description of multiple satellites},\
  }\href {https://doi.org/10.1103/PhysRevLett.107.166401} {\bibfield  {journal}
  {\bibinfo  {journal} {Phys. Rev. Lett.}\ }\textbf {\bibinfo {volume} {107}},\
  \bibinfo {pages} {166401} (\bibinfo {year} {2011})}\BibitemShut {NoStop}%
\bibitem [{\citenamefont {Riva}\ \emph {et~al.}(2023)\citenamefont {Riva},
  \citenamefont {Romaniello},\ and\ \citenamefont {Berger}}]{riva_prl}%
  \BibitemOpen
  \bibfield  {author} {\bibinfo {author} {\bibfnamefont {G.}~\bibnamefont
  {Riva}}, \bibinfo {author} {\bibfnamefont {P.}~\bibnamefont {Romaniello}},\
  and\ \bibinfo {author} {\bibfnamefont {J.~A.}\ \bibnamefont {Berger}},\
  }\bibfield  {title} {\bibinfo {title} {Multichannel dyson equation: Coupling
  many-body green's functions},\ }\href
  {https://doi.org/10.1103/PhysRevLett.131.216401} {\bibfield  {journal}
  {\bibinfo  {journal} {Phys. Rev. Lett.}\ }\textbf {\bibinfo {volume} {131}},\
  \bibinfo {pages} {216401} (\bibinfo {year} {2023})}\BibitemShut {NoStop}%
\bibitem [{\citenamefont {Mejuto-Zaera}\ \emph {et~al.}(2021)\citenamefont
  {Mejuto-Zaera}, \citenamefont {Weng}, \citenamefont {Romanova}, \citenamefont
  {Cotton}, \citenamefont {Whaley}, \citenamefont {Tubman},\ and\ \citenamefont
  {Vlček}}]{Mejuto_2021}%
  \BibitemOpen
  \bibfield  {author} {\bibinfo {author} {\bibfnamefont {C.}~\bibnamefont
  {Mejuto-Zaera}}, \bibinfo {author} {\bibfnamefont {G.}~\bibnamefont {Weng}},
  \bibinfo {author} {\bibfnamefont {M.}~\bibnamefont {Romanova}}, \bibinfo
  {author} {\bibfnamefont {S.~J.}\ \bibnamefont {Cotton}}, \bibinfo {author}
  {\bibfnamefont {K.~B.}\ \bibnamefont {Whaley}}, \bibinfo {author}
  {\bibfnamefont {N.~M.}\ \bibnamefont {Tubman}},\ and\ \bibinfo {author}
  {\bibfnamefont {V.}~\bibnamefont {Vlček}},\ }\bibfield  {title} {\bibinfo
  {title} {Are multi-quasiparticle interactions important in molecular
  ionization?},\ }\href {https://doi.org/10.1063/5.0044060} {\bibfield
  {journal} {\bibinfo  {journal} {The Journal of Chemical Physics}\ }\textbf
  {\bibinfo {volume} {154}},\ \bibinfo {pages} {121101} (\bibinfo {year}
  {2021})}\BibitemShut {NoStop}%
\bibitem [{\citenamefont {Lani}\ \emph {et~al.}(2012)\citenamefont {Lani},
  \citenamefont {Romaniello},\ and\ \citenamefont {Reining}}]{Lan12}%
  \BibitemOpen
  \bibfield  {author} {\bibinfo {author} {\bibfnamefont {G.}~\bibnamefont
  {Lani}}, \bibinfo {author} {\bibfnamefont {P.}~\bibnamefont {Romaniello}},\
  and\ \bibinfo {author} {\bibfnamefont {L.}~\bibnamefont {Reining}},\
  }\bibfield  {title} {\bibinfo {title} {Approximations for many-body green's
  functions: insights from the fundamental equations},\ }\href
  {https://doi.org/10.1088/1367-2630/14/1/013056} {\bibfield  {journal}
  {\bibinfo  {journal} {New Journal of Physics}\ }\textbf {\bibinfo {volume}
  {14}},\ \bibinfo {pages} {013056} (\bibinfo {year} {2012})}\BibitemShut
  {NoStop}%
\bibitem [{\citenamefont {Berger}\ \emph {et~al.}(2014)\citenamefont {Berger},
  \citenamefont {Romaniello}, \citenamefont {Tandetzky}, \citenamefont
  {Mendoza}, \citenamefont {Brouder},\ and\ \citenamefont {Reining}}]{Ber14}%
  \BibitemOpen
  \bibfield  {author} {\bibinfo {author} {\bibfnamefont {J.~A.}\ \bibnamefont
  {Berger}}, \bibinfo {author} {\bibfnamefont {P.}~\bibnamefont {Romaniello}},
  \bibinfo {author} {\bibfnamefont {F.}~\bibnamefont {Tandetzky}}, \bibinfo
  {author} {\bibfnamefont {B.~S.}\ \bibnamefont {Mendoza}}, \bibinfo {author}
  {\bibfnamefont {C.}~\bibnamefont {Brouder}},\ and\ \bibinfo {author}
  {\bibfnamefont {L.}~\bibnamefont {Reining}},\ }\bibfield  {title} {\bibinfo
  {title} {Solution to the many-body problem in one point},\ }\href
  {https://doi.org/10.1088/1367-2630/16/11/113025} {\bibfield  {journal}
  {\bibinfo  {journal} {New Journal of Physics}\ }\textbf {\bibinfo {volume}
  {16}},\ \bibinfo {pages} {113025} (\bibinfo {year} {2014})}\BibitemShut
  {NoStop}%
\bibitem [{\citenamefont {Stan}\ \emph {et~al.}(2015)\citenamefont {Stan},
  \citenamefont {Romaniello}, \citenamefont {Rigamonti}, \citenamefont
  {Reining},\ and\ \citenamefont {Berger}}]{Sta15}%
  \BibitemOpen
  \bibfield  {author} {\bibinfo {author} {\bibfnamefont {A.}~\bibnamefont
  {Stan}}, \bibinfo {author} {\bibfnamefont {P.}~\bibnamefont {Romaniello}},
  \bibinfo {author} {\bibfnamefont {S.}~\bibnamefont {Rigamonti}}, \bibinfo
  {author} {\bibfnamefont {L.}~\bibnamefont {Reining}},\ and\ \bibinfo {author}
  {\bibfnamefont {J.~A.}\ \bibnamefont {Berger}},\ }\bibfield  {title}
  {\bibinfo {title} {Unphysical and physical solutions in many-body theories:
  from weak to strong correlation},\ }\href
  {https://doi.org/10.1088/1367-2630/17/9/093045} {\bibfield  {journal}
  {\bibinfo  {journal} {New Journal of Physics}\ }\textbf {\bibinfo {volume}
  {17}},\ \bibinfo {pages} {093045} (\bibinfo {year} {2015})}\BibitemShut
  {NoStop}%
\bibitem [{\citenamefont {Loos}\ \emph {et~al.}(2018)\citenamefont {Loos},
  \citenamefont {Romaniello},\ and\ \citenamefont {Berger}}]{Loo18}%
  \BibitemOpen
  \bibfield  {author} {\bibinfo {author} {\bibfnamefont {P.-F.}\ \bibnamefont
  {Loos}}, \bibinfo {author} {\bibfnamefont {P.}~\bibnamefont {Romaniello}},\
  and\ \bibinfo {author} {\bibfnamefont {J.~A.}\ \bibnamefont {Berger}},\
  }\bibfield  {title} {\bibinfo {title} {Green functions and self-consistency:
  Insights from the spherium model},\ }\href
  {https://doi.org/10.1021/acs.jctc.8b00260} {\bibfield  {journal} {\bibinfo
  {journal} {Journal of Chemical Theory and Computation}\ }\textbf {\bibinfo
  {volume} {14}},\ \bibinfo {pages} {3071} (\bibinfo {year} {2018})},\ \bibinfo
  {note} {pMID: 29746773}\BibitemShut {NoStop}%
\bibitem [{\citenamefont {Véril}\ \emph {et~al.}(2018)\citenamefont {Véril},
  \citenamefont {Romaniello}, \citenamefont {Berger},\ and\ \citenamefont
  {Loos}}]{Vér18}%
  \BibitemOpen
  \bibfield  {author} {\bibinfo {author} {\bibfnamefont {M.}~\bibnamefont
  {Véril}}, \bibinfo {author} {\bibfnamefont {P.}~\bibnamefont {Romaniello}},
  \bibinfo {author} {\bibfnamefont {J.~A.}\ \bibnamefont {Berger}},\ and\
  \bibinfo {author} {\bibfnamefont {P.-F.}\ \bibnamefont {Loos}},\ }\bibfield
  {title} {\bibinfo {title} {Unphysical discontinuities in gw methods},\ }\href
  {https://doi.org/10.1021/acs.jctc.8b00745} {\bibfield  {journal} {\bibinfo
  {journal} {Journal of Chemical Theory and Computation}\ }\textbf {\bibinfo
  {volume} {14}},\ \bibinfo {pages} {5220} (\bibinfo {year}
  {2018})}\BibitemShut {NoStop}%
\bibitem [{\citenamefont {Tarantino}\ \emph {et~al.}(2017)\citenamefont
  {Tarantino}, \citenamefont {Romaniello}, \citenamefont {Berger},\ and\
  \citenamefont {Reining}}]{Tar17}%
  \BibitemOpen
  \bibfield  {author} {\bibinfo {author} {\bibfnamefont {W.}~\bibnamefont
  {Tarantino}}, \bibinfo {author} {\bibfnamefont {P.}~\bibnamefont
  {Romaniello}}, \bibinfo {author} {\bibfnamefont {J.~A.}\ \bibnamefont
  {Berger}},\ and\ \bibinfo {author} {\bibfnamefont {L.}~\bibnamefont
  {Reining}},\ }\bibfield  {title} {\bibinfo {title} {Self-consistent dyson
  equation and self-energy functionals: An analysis and illustration on the
  example of the hubbard atom},\ }\href
  {https://doi.org/10.1103/PhysRevB.96.045124} {\bibfield  {journal} {\bibinfo
  {journal} {Phys. Rev. B}\ }\textbf {\bibinfo {volume} {96}},\ \bibinfo
  {pages} {045124} (\bibinfo {year} {2017})}\BibitemShut {NoStop}%
\bibitem [{\citenamefont {Lischner}\ \emph {et~al.}(2013)\citenamefont
  {Lischner}, \citenamefont {Vigil-Fowler},\ and\ \citenamefont
  {Louie}}]{Lischner_2013}%
  \BibitemOpen
  \bibfield  {author} {\bibinfo {author} {\bibfnamefont {J.}~\bibnamefont
  {Lischner}}, \bibinfo {author} {\bibfnamefont {D.}~\bibnamefont
  {Vigil-Fowler}},\ and\ \bibinfo {author} {\bibfnamefont {S.~G.}\ \bibnamefont
  {Louie}},\ }\bibfield  {title} {\bibinfo {title} {Physical origin of
  satellites in photoemission of doped graphene: An ab initio {$GW$} plus
  cumulant study},\ }\href {https://doi.org/10.1103/PhysRevLett.110.146801}
  {\bibfield  {journal} {\bibinfo  {journal} {Phys. Rev. Lett.}\ }\textbf
  {\bibinfo {volume} {110}},\ \bibinfo {pages} {146801} (\bibinfo {year}
  {2013})}\BibitemShut {NoStop}%
\bibitem [{\citenamefont {Zhou}\ \emph {et~al.}(2015)\citenamefont {Zhou},
  \citenamefont {Kas}, \citenamefont {Sponza}, \citenamefont {Reshetnyak},
  \citenamefont {Guzzo}, \citenamefont {Giorgetti}, \citenamefont {Gatti},
  \citenamefont {Sottile}, \citenamefont {Rehr},\ and\ \citenamefont
  {Reining}}]{Zhou_2015}%
  \BibitemOpen
  \bibfield  {author} {\bibinfo {author} {\bibfnamefont {J.~S.}\ \bibnamefont
  {Zhou}}, \bibinfo {author} {\bibfnamefont {J.~J.}\ \bibnamefont {Kas}},
  \bibinfo {author} {\bibfnamefont {L.}~\bibnamefont {Sponza}}, \bibinfo
  {author} {\bibfnamefont {I.}~\bibnamefont {Reshetnyak}}, \bibinfo {author}
  {\bibfnamefont {M.}~\bibnamefont {Guzzo}}, \bibinfo {author} {\bibfnamefont
  {C.}~\bibnamefont {Giorgetti}}, \bibinfo {author} {\bibfnamefont
  {M.}~\bibnamefont {Gatti}}, \bibinfo {author} {\bibfnamefont
  {F.}~\bibnamefont {Sottile}}, \bibinfo {author} {\bibfnamefont {J.~J.}\
  \bibnamefont {Rehr}},\ and\ \bibinfo {author} {\bibfnamefont
  {L.}~\bibnamefont {Reining}},\ }\bibfield  {title} {\bibinfo {title}
  {Dynamical effects in electron spectroscopy},\ }\href
  {https://doi.org/10.1063/1.4934965} {\bibfield  {journal} {\bibinfo
  {journal} {J. Chem. Phys.}\ }\textbf {\bibinfo {volume} {143}},\ \bibinfo
  {pages} {184109} (\bibinfo {year} {2015})}\BibitemShut {NoStop}%
\bibitem [{\citenamefont {Gumhalter}\ \emph {et~al.}(2016)\citenamefont
  {Gumhalter}, \citenamefont {Kova\ifmmode~\check{c}\else \v{c}\fi{}},
  \citenamefont {Caruso}, \citenamefont {Lambert},\ and\ \citenamefont
  {Giustino}}]{Gumhalter_2016}%
  \BibitemOpen
  \bibfield  {author} {\bibinfo {author} {\bibfnamefont {B.}~\bibnamefont
  {Gumhalter}}, \bibinfo {author} {\bibfnamefont {V.}~\bibnamefont
  {Kova\ifmmode~\check{c}\else \v{c}\fi{}}}, \bibinfo {author} {\bibfnamefont
  {F.}~\bibnamefont {Caruso}}, \bibinfo {author} {\bibfnamefont
  {H.}~\bibnamefont {Lambert}},\ and\ \bibinfo {author} {\bibfnamefont
  {F.}~\bibnamefont {Giustino}},\ }\bibfield  {title} {\bibinfo {title} {On the
  combined use of gw approximation and cumulant expansion in the calculations
  of quasiparticle spectra: The paradigm of si valence bands},\ }\href
  {https://doi.org/10.1103/PhysRevB.94.035103} {\bibfield  {journal} {\bibinfo
  {journal} {Phys. Rev. B}\ }\textbf {\bibinfo {volume} {94}},\ \bibinfo
  {pages} {035103} (\bibinfo {year} {2016})}\BibitemShut {NoStop}%
\bibitem [{\citenamefont {Kockläuner}\ and\ \citenamefont
  {Golze}(2025)}]{Kocklauner_2025}%
  \BibitemOpen
  \bibfield  {author} {\bibinfo {author} {\bibfnamefont {J.}~\bibnamefont
  {Kockläuner}}\ and\ \bibinfo {author} {\bibfnamefont {D.}~\bibnamefont
  {Golze}},\ }\bibfield  {title} {\bibinfo {title} {Gw plus cumulant approach
  for predicting core-level shakeup satellites in large molecules},\ }\href
  {https://doi.org/10.1021/acs.jctc.4c01754} {\bibfield  {journal} {\bibinfo
  {journal} {Journal of Chemical Theory and Computation}\ }\textbf {\bibinfo
  {volume} {21}},\ \bibinfo {pages} {3101} (\bibinfo {year} {2025})},\ \bibinfo
  {note} {pMID: 40029694}\BibitemShut {NoStop}%
\bibitem [{\citenamefont {Brisk}\ and\ \citenamefont
  {Baker}(1975)}]{Brisk_1975}%
  \BibitemOpen
  \bibfield  {author} {\bibinfo {author} {\bibfnamefont {M.~A.}\ \bibnamefont
  {Brisk}}\ and\ \bibinfo {author} {\bibfnamefont {A.}~\bibnamefont {Baker}},\
  }\bibfield  {title} {\bibinfo {title} {Shake-up satellites in x-ray
  photoelectron spectroscopy},\ }\href
  {https://doi.org/https://doi.org/10.1016/0368-2048(75)80061-2} {\bibfield
  {journal} {\bibinfo  {journal} {Journal of Electron Spectroscopy and Related
  Phenomena}\ }\textbf {\bibinfo {volume} {7}},\ \bibinfo {pages} {197}
  (\bibinfo {year} {1975})}\BibitemShut {NoStop}%
\bibitem [{\citenamefont {Arneberg}\ \emph {et~al.}(1982)\citenamefont
  {Arneberg}, \citenamefont {Müller},\ and\ \citenamefont
  {Manne}}]{Arneberg_1982}%
  \BibitemOpen
  \bibfield  {author} {\bibinfo {author} {\bibfnamefont {R.}~\bibnamefont
  {Arneberg}}, \bibinfo {author} {\bibfnamefont {J.}~\bibnamefont {Müller}},\
  and\ \bibinfo {author} {\bibfnamefont {R.}~\bibnamefont {Manne}},\ }\bibfield
   {title} {\bibinfo {title} {Configuration interaction calculations of
  satellite structure in photoelectron spectra of h2o},\ }\href
  {https://doi.org/https://doi.org/10.1016/0301-0104(82)87091-2} {\bibfield
  {journal} {\bibinfo  {journal} {Chemical Physics}\ }\textbf {\bibinfo
  {volume} {64}},\ \bibinfo {pages} {249} (\bibinfo {year} {1982})}\BibitemShut
  {NoStop}%
\bibitem [{\citenamefont {Chatterjee}\ and\ \citenamefont
  {Sokolov}(2019)}]{Chatterjee_2019}%
  \BibitemOpen
  \bibfield  {author} {\bibinfo {author} {\bibfnamefont {K.}~\bibnamefont
  {Chatterjee}}\ and\ \bibinfo {author} {\bibfnamefont {A.~Y.}\ \bibnamefont
  {Sokolov}},\ }\bibfield  {title} {\bibinfo {title} {Second-order
  multireference algebraic diagrammatic construction theory for photoelectron
  spectra of strongly correlated systems},\ }\href
  {https://doi.org/10.1021/acs.jctc.9b00528} {\bibfield  {journal} {\bibinfo
  {journal} {Journal of Chemical Theory and Computation}\ }\textbf {\bibinfo
  {volume} {15}},\ \bibinfo {pages} {5908} (\bibinfo {year} {2019})},\ \bibinfo
  {note} {pMID: 31509706}\BibitemShut {NoStop}%
\bibitem [{\citenamefont {Nakatsuji}(1991)}]{Nakatsuji_1991}%
  \BibitemOpen
  \bibfield  {author} {\bibinfo {author} {\bibfnamefont {H.}~\bibnamefont
  {Nakatsuji}},\ }\bibfield  {title} {\bibinfo {title} {Description of two- and
  many-electron processes by the sac-ci method},\ }\href
  {https://doi.org/https://doi.org/10.1016/0009-2614(91)85040-4} {\bibfield
  {journal} {\bibinfo  {journal} {Chemical Physics Letters}\ }\textbf {\bibinfo
  {volume} {177}},\ \bibinfo {pages} {331} (\bibinfo {year}
  {1991})}\BibitemShut {NoStop}%
\bibitem [{\citenamefont {Kuramoto}\ \emph {et~al.}(2004)\citenamefont
  {Kuramoto}, \citenamefont {Ehara},\ and\ \citenamefont
  {Nakatsuji}}]{Kuramoto_2004}%
  \BibitemOpen
  \bibfield  {author} {\bibinfo {author} {\bibfnamefont {K.}~\bibnamefont
  {Kuramoto}}, \bibinfo {author} {\bibfnamefont {M.}~\bibnamefont {Ehara}},\
  and\ \bibinfo {author} {\bibfnamefont {H.}~\bibnamefont {Nakatsuji}},\
  }\bibfield  {title} {\bibinfo {title} {Theoretical fine spectroscopy with
  symmetry adapted cluster-configuration interaction general-r method:
  First-row k-shell ionizations and their satellites},\ }\href
  {https://doi.org/10.1063/1.1824899} {\bibfield  {journal} {\bibinfo
  {journal} {J. Chem. Phys.}\ }\textbf {\bibinfo {volume} {122}},\ \bibinfo
  {pages} {014304} (\bibinfo {year} {2004})}\BibitemShut {NoStop}%
\bibitem [{\citenamefont {Marie}\ and\ \citenamefont
  {Loos}(2024)}]{Marie_2024}%
  \BibitemOpen
  \bibfield  {author} {\bibinfo {author} {\bibfnamefont {A.}~\bibnamefont
  {Marie}}\ and\ \bibinfo {author} {\bibfnamefont {P.-F.}\ \bibnamefont
  {Loos}},\ }\bibfield  {title} {\bibinfo {title} {Reference energies for
  valence ionizations and satellite transitions},\ }\href
  {https://doi.org/10.1021/acs.jctc.4c00216} {\bibfield  {journal} {\bibinfo
  {journal} {Journal of Chemical Theory and Computation}\ }\textbf {\bibinfo
  {volume} {20}},\ \bibinfo {pages} {4751} (\bibinfo {year} {2024})},\ \bibinfo
  {note} {pMID: 38776293}\BibitemShut {NoStop}%
\bibitem [{\citenamefont {Lisini}\ and\ \citenamefont
  {Decleva}(1992)}]{Lisini_1992}%
  \BibitemOpen
  \bibfield  {author} {\bibinfo {author} {\bibfnamefont {A.}~\bibnamefont
  {Lisini}}\ and\ \bibinfo {author} {\bibfnamefont {P.}~\bibnamefont
  {Decleva}},\ }\bibfield  {title} {\bibinfo {title} {Calculation of dynamical
  correlation effects by quasidegenerate perturbation theory. an application to
  photoionization spectra},\ }\href
  {https://doi.org/https://doi.org/10.1016/0301-0104(92)80103-3} {\bibfield
  {journal} {\bibinfo  {journal} {Chemical Physics}\ }\textbf {\bibinfo
  {volume} {168}},\ \bibinfo {pages} {1} (\bibinfo {year} {1992})}\BibitemShut
  {NoStop}%
\bibitem [{\citenamefont {Lisini}\ \emph {et~al.}(1993)\citenamefont {Lisini},
  \citenamefont {Decleva},\ and\ \citenamefont {Fronzoni}}]{Lisini_1993}%
  \BibitemOpen
  \bibfield  {author} {\bibinfo {author} {\bibfnamefont {A.}~\bibnamefont
  {Lisini}}, \bibinfo {author} {\bibfnamefont {P.}~\bibnamefont {Decleva}},\
  and\ \bibinfo {author} {\bibfnamefont {G.}~\bibnamefont {Fronzoni}},\
  }\bibfield  {title} {\bibinfo {title} {Quasidegenerate perturbation theory
  applied to the calculation of excitation spectra},\ }\href
  {https://doi.org/https://doi.org/10.1016/0301-0104(93)85140-4} {\bibfield
  {journal} {\bibinfo  {journal} {Chemical Physics}\ }\textbf {\bibinfo
  {volume} {171}},\ \bibinfo {pages} {159} (\bibinfo {year}
  {1993})}\BibitemShut {NoStop}%
\bibitem [{\citenamefont {Fronzoni}\ \emph {et~al.}(1999)\citenamefont
  {Fronzoni}, \citenamefont {Alti},\ and\ \citenamefont
  {Decleva}}]{Fronzoni_1999}%
  \BibitemOpen
  \bibfield  {author} {\bibinfo {author} {\bibfnamefont {G.}~\bibnamefont
  {Fronzoni}}, \bibinfo {author} {\bibfnamefont {G.~D.}\ \bibnamefont {Alti}},\
  and\ \bibinfo {author} {\bibfnamefont {P.}~\bibnamefont {Decleva}},\
  }\bibfield  {title} {\bibinfo {title} {Theoretical c1s and o1s core shake-up
  spectra of co by highly correlated qdptci approach},\ }\href
  {https://doi.org/10.1088/0953-4075/32/22/313} {\bibfield  {journal} {\bibinfo
   {journal} {Journal of Physics B: Atomic, Molecular and Optical Physics}\
  }\textbf {\bibinfo {volume} {32}},\ \bibinfo {pages} {5357} (\bibinfo {year}
  {1999})}\BibitemShut {NoStop}%
\bibitem [{\citenamefont {Nooijen}\ and\ \citenamefont
  {Snijders}(1992)}]{Nooijen_1992}%
  \BibitemOpen
  \bibfield  {author} {\bibinfo {author} {\bibfnamefont {M.}~\bibnamefont
  {Nooijen}}\ and\ \bibinfo {author} {\bibfnamefont {J.~G.}\ \bibnamefont
  {Snijders}},\ }\bibfield  {title} {\bibinfo {title} {Coupled cluster approach
  to the single-particle green's function},\ }\href
  {https://doi.org/https://doi.org/10.1002/qua.560440808} {\bibfield  {journal}
  {\bibinfo  {journal} {Int. J. Quantum. Chem.}\ }\textbf {\bibinfo {volume}
  {44}},\ \bibinfo {pages} {55} (\bibinfo {year} {1992})}\BibitemShut {NoStop}%
\bibitem [{\citenamefont {Nooijen}\ and\ \citenamefont
  {Snijders}(1993)}]{Nooijen_1993}%
  \BibitemOpen
  \bibfield  {author} {\bibinfo {author} {\bibfnamefont {M.}~\bibnamefont
  {Nooijen}}\ and\ \bibinfo {author} {\bibfnamefont {J.~G.}\ \bibnamefont
  {Snijders}},\ }\bibfield  {title} {\bibinfo {title} {Coupled cluster green's
  function method: Working equations and applications},\ }\href
  {https://doi.org/https://doi.org/10.1002/qua.560480103} {\bibfield  {journal}
  {\bibinfo  {journal} {Int. J. Quantum. Chem.}\ }\textbf {\bibinfo {volume}
  {48}},\ \bibinfo {pages} {15} (\bibinfo {year} {1993})}\BibitemShut {NoStop}%
\bibitem [{\citenamefont {Nooijen}\ and\ \citenamefont
  {Snijders}(1995)}]{Nooijen_1995}%
  \BibitemOpen
  \bibfield  {author} {\bibinfo {author} {\bibfnamefont {M.}~\bibnamefont
  {Nooijen}}\ and\ \bibinfo {author} {\bibfnamefont {J.~G.}\ \bibnamefont
  {Snijders}},\ }\bibfield  {title} {\bibinfo {title} {Second order many‐body
  perturbation approximations to the coupled cluster green’s function},\
  }\href {https://doi.org/10.1063/1.468900} {\bibfield  {journal} {\bibinfo
  {journal} {J. Chem. Phys.}\ }\textbf {\bibinfo {volume} {102}},\ \bibinfo
  {pages} {1681} (\bibinfo {year} {1995})}\BibitemShut {NoStop}%
\bibitem [{\citenamefont {Rehr}\ \emph {et~al.}(2020)\citenamefont {Rehr},
  \citenamefont {Vila}, \citenamefont {Kas}, \citenamefont {Hirshberg},
  \citenamefont {Kowalski},\ and\ \citenamefont {Peng}}]{Rehr_2020}%
  \BibitemOpen
  \bibfield  {author} {\bibinfo {author} {\bibfnamefont {J.~J.}\ \bibnamefont
  {Rehr}}, \bibinfo {author} {\bibfnamefont {F.~D.}\ \bibnamefont {Vila}},
  \bibinfo {author} {\bibfnamefont {J.~J.}\ \bibnamefont {Kas}}, \bibinfo
  {author} {\bibfnamefont {N.~Y.}\ \bibnamefont {Hirshberg}}, \bibinfo {author}
  {\bibfnamefont {K.}~\bibnamefont {Kowalski}},\ and\ \bibinfo {author}
  {\bibfnamefont {B.}~\bibnamefont {Peng}},\ }\bibfield  {title} {\bibinfo
  {title} {Equation of motion coupled-cluster cumulant approach for intrinsic
  losses in x-ray spectra},\ }\href {https://doi.org/10.1063/5.0004865}
  {\bibfield  {journal} {\bibinfo  {journal} {J. Chem. Phys.}\ }\textbf
  {\bibinfo {volume} {152}},\ \bibinfo {pages} {174113} (\bibinfo {year}
  {2020})}\BibitemShut {NoStop}%
\bibitem [{\citenamefont {Vila}\ \emph {et~al.}(2020)\citenamefont {Vila},
  \citenamefont {Rehr}, \citenamefont {Kas}, \citenamefont {Kowalski},\ and\
  \citenamefont {Peng}}]{Vila_2020}%
  \BibitemOpen
  \bibfield  {author} {\bibinfo {author} {\bibfnamefont {F.~D.}\ \bibnamefont
  {Vila}}, \bibinfo {author} {\bibfnamefont {J.~J.}\ \bibnamefont {Rehr}},
  \bibinfo {author} {\bibfnamefont {J.~J.}\ \bibnamefont {Kas}}, \bibinfo
  {author} {\bibfnamefont {K.}~\bibnamefont {Kowalski}},\ and\ \bibinfo
  {author} {\bibfnamefont {B.}~\bibnamefont {Peng}},\ }\bibfield  {title}
  {\bibinfo {title} {Real-time coupled-cluster approach for the cumulant
  green’s function},\ }\href {https://doi.org/10.1021/acs.jctc.0c00639}
  {\bibfield  {journal} {\bibinfo  {journal} {Journal of Chemical Theory and
  Computation}\ }\textbf {\bibinfo {volume} {16}},\ \bibinfo {pages} {6983}
  (\bibinfo {year} {2020})},\ \bibinfo {note} {pMID: 33108872}\BibitemShut
  {NoStop}%
\bibitem [{\citenamefont {Vila}\ \emph {et~al.}(2021)\citenamefont {Vila},
  \citenamefont {Kas}, \citenamefont {Rehr}, \citenamefont {Kowalski},\ and\
  \citenamefont {Peng}}]{Vila_2021}%
  \BibitemOpen
  \bibfield  {author} {\bibinfo {author} {\bibfnamefont {F.~D.}\ \bibnamefont
  {Vila}}, \bibinfo {author} {\bibfnamefont {J.~J.}\ \bibnamefont {Kas}},
  \bibinfo {author} {\bibfnamefont {J.~J.}\ \bibnamefont {Rehr}}, \bibinfo
  {author} {\bibfnamefont {K.}~\bibnamefont {Kowalski}},\ and\ \bibinfo
  {author} {\bibfnamefont {B.}~\bibnamefont {Peng}},\ }\bibfield  {title}
  {\bibinfo {title} {Equation-of-motion coupled-cluster cumulant green’s
  function for excited states and x-ray spectra},\ }\bibfield  {journal}
  {\bibinfo  {journal} {Frontiers in Chemistry}\ }\textbf {\bibinfo {volume}
  {Volume 9 - 2021}},\ \href {https://doi.org/10.3389/fchem.2021.734945}
  {10.3389/fchem.2021.734945} (\bibinfo {year} {2021})\BibitemShut {NoStop}%
\bibitem [{\citenamefont {Vila}\ \emph {et~al.}(2022)\citenamefont {Vila},
  \citenamefont {Kowalski}, \citenamefont {Peng}, \citenamefont {Kas},\ and\
  \citenamefont {Rehr}}]{Vila_2022}%
  \BibitemOpen
  \bibfield  {author} {\bibinfo {author} {\bibfnamefont {F.~D.}\ \bibnamefont
  {Vila}}, \bibinfo {author} {\bibfnamefont {K.}~\bibnamefont {Kowalski}},
  \bibinfo {author} {\bibfnamefont {B.}~\bibnamefont {Peng}}, \bibinfo {author}
  {\bibfnamefont {J.~J.}\ \bibnamefont {Kas}},\ and\ \bibinfo {author}
  {\bibfnamefont {J.~J.}\ \bibnamefont {Rehr}},\ }\bibfield  {title} {\bibinfo
  {title} {Real-time equation-of-motion ccsd cumulant green’s function},\
  }\href {https://doi.org/10.1021/acs.jctc.1c01179} {\bibfield  {journal}
  {\bibinfo  {journal} {Journal of Chemical Theory and Computation}\ }\textbf
  {\bibinfo {volume} {18}},\ \bibinfo {pages} {1799} (\bibinfo {year}
  {2022})},\ \bibinfo {note} {pMID: 35157796}\BibitemShut {NoStop}%
\bibitem [{\citenamefont {Riva}\ \emph {et~al.}(2024)\citenamefont {Riva},
  \citenamefont {Romaniello},\ and\ \citenamefont {Berger}}]{riva_prb}%
  \BibitemOpen
  \bibfield  {author} {\bibinfo {author} {\bibfnamefont {G.}~\bibnamefont
  {Riva}}, \bibinfo {author} {\bibfnamefont {P.}~\bibnamefont {Romaniello}},\
  and\ \bibinfo {author} {\bibfnamefont {J.~A.}\ \bibnamefont {Berger}},\
  }\bibfield  {title} {\bibinfo {title} {Derivation and analysis of the
  multichannel dyson equation},\ }\href
  {https://doi.org/10.1103/PhysRevB.110.115140} {\bibfield  {journal} {\bibinfo
   {journal} {Phys. Rev. B}\ }\textbf {\bibinfo {volume} {110}},\ \bibinfo
  {pages} {115140} (\bibinfo {year} {2024})}\BibitemShut {NoStop}%
\bibitem [{\citenamefont {Riva}\ \emph {et~al.}(2025)\citenamefont {Riva},
  \citenamefont {Fischer}, \citenamefont {Paggi}, \citenamefont {Berger},\ and\
  \citenamefont {Romaniello}}]{riva_prb_25}%
  \BibitemOpen
  \bibfield  {author} {\bibinfo {author} {\bibfnamefont {G.}~\bibnamefont
  {Riva}}, \bibinfo {author} {\bibfnamefont {T.}~\bibnamefont {Fischer}},
  \bibinfo {author} {\bibfnamefont {S.}~\bibnamefont {Paggi}}, \bibinfo
  {author} {\bibfnamefont {J.~A.}\ \bibnamefont {Berger}},\ and\ \bibinfo
  {author} {\bibfnamefont {P.}~\bibnamefont {Romaniello}},\ }\bibfield  {title}
  {\bibinfo {title} {Multichannel dyson equations for even- and odd-order
  green's functions: Application to double excitations},\ }\href
  {https://doi.org/10.1103/PhysRevB.111.195133} {\bibfield  {journal} {\bibinfo
   {journal} {Phys. Rev. B}\ }\textbf {\bibinfo {volume} {111}},\ \bibinfo
  {pages} {195133} (\bibinfo {year} {2025})}\BibitemShut {NoStop}%
\bibitem [{\citenamefont {Paggi}\ \emph {et~al.}(2025)\citenamefont {Paggi},
  \citenamefont {Berger},\ and\ \citenamefont {Romaniello}}]{paggi_25}%
  \BibitemOpen
  \bibfield  {author} {\bibinfo {author} {\bibfnamefont {S.}~\bibnamefont
  {Paggi}}, \bibinfo {author} {\bibfnamefont {J.~A.}\ \bibnamefont {Berger}},\
  and\ \bibinfo {author} {\bibfnamefont {P.}~\bibnamefont {Romaniello}},\
  }\bibfield  {title} {\bibinfo {title} {Ground and excited-state properties of
  the extended hubbard dimer from the multichannel dyson equation},\ }\href
  {https://doi.org/10.1063/5.0291280} {\bibfield  {journal} {\bibinfo
  {journal} {J. Chem. Phys.}\ }\textbf {\bibinfo {volume} {163}},\ \bibinfo
  {pages} {154109} (\bibinfo {year} {2025})}\BibitemShut {NoStop}%
\bibitem [{\citenamefont {Selli\'e}\ \emph {et~al.}(2026)\citenamefont
  {Selli\'e}, \citenamefont {Berger},\ and\ \citenamefont
  {Romaniello}}]{Sellie_2026}%
  \BibitemOpen
  \bibfield  {author} {\bibinfo {author} {\bibfnamefont {P.}~\bibnamefont
  {Selli\'e}}, \bibinfo {author} {\bibfnamefont {J.~A.}\ \bibnamefont
  {Berger}},\ and\ \bibinfo {author} {\bibfnamefont {P.}~\bibnamefont
  {Romaniello}},\ }\bibfield  {title} {\bibinfo {title} {{Multichannel Dyson
  equation for double ionization spectroscopies}},\ }\href
  {https://doi.org/10.1103/jcfy-vdm8} {\bibfield  {journal} {\bibinfo
  {journal} {Phys. Rev. B}\ }\textbf {\bibinfo {volume} {114}},\ \bibinfo
  {pages} {125117} (\bibinfo {year} {2026})}\BibitemShut {NoStop}%
\bibitem [{\citenamefont {Riva}\ \emph {et~al.}(2022)\citenamefont {Riva},
  \citenamefont {Audinet}, \citenamefont {Vladaj}, \citenamefont {Romaniello},\
  and\ \citenamefont {Berger}}]{Riv22}%
  \BibitemOpen
  \bibfield  {author} {\bibinfo {author} {\bibfnamefont {G.}~\bibnamefont
  {Riva}}, \bibinfo {author} {\bibfnamefont {T.}~\bibnamefont {Audinet}},
  \bibinfo {author} {\bibfnamefont {M.}~\bibnamefont {Vladaj}}, \bibinfo
  {author} {\bibfnamefont {P.}~\bibnamefont {Romaniello}},\ and\ \bibinfo
  {author} {\bibfnamefont {J.~A.}\ \bibnamefont {Berger}},\ }\bibfield  {title}
  {\bibinfo {title} {{Photoemission spectral functions from the three-body
  Green's function}},\ }\href {https://doi.org/10.21468/SciPostPhys.12.3.093}
  {\bibfield  {journal} {\bibinfo  {journal} {SciPost Phys.}\ }\textbf
  {\bibinfo {volume} {12}},\ \bibinfo {pages} {093} (\bibinfo {year}
  {2022})}\BibitemShut {NoStop}%
\bibitem [{\citenamefont {Romaniello}\ and\ \citenamefont
  {Berger}(2026)}]{Romaniello_2026}%
  \BibitemOpen
  \bibfield  {author} {\bibinfo {author} {\bibfnamefont {P.}~\bibnamefont
  {Romaniello}}\ and\ \bibinfo {author} {\bibfnamefont {J.~A.}\ \bibnamefont
  {Berger}},\ }\href {https://arxiv.org/abs/2603.27329} {\bibinfo {title}
  {Direct and inverse photoemission spectra from the screened multichannel
  dyson equation}} (\bibinfo {year} {2026}),\ \Eprint
  {https://arxiv.org/abs/2603.27329} {arXiv:2603.27329 [cond-mat.other]}
  \BibitemShut {NoStop}%
\bibitem [{\citenamefont {Schirmer}\ \emph {et~al.}(1983)\citenamefont
  {Schirmer}, \citenamefont {Cederbaum},\ and\ \citenamefont {Walter}}]{Sch83}%
  \BibitemOpen
  \bibfield  {author} {\bibinfo {author} {\bibfnamefont {J.}~\bibnamefont
  {Schirmer}}, \bibinfo {author} {\bibfnamefont {L.~S.}\ \bibnamefont
  {Cederbaum}},\ and\ \bibinfo {author} {\bibfnamefont {O.}~\bibnamefont
  {Walter}},\ }\bibfield  {title} {\bibinfo {title} {{New approach to the
  one-particle Green's function for finite Fermi systems}},\ }\href
  {https://doi.org/10.1103/PhysRevA.28.1237} {\bibfield  {journal} {\bibinfo
  {journal} {Phys. Rev. A}\ }\textbf {\bibinfo {volume} {28}},\ \bibinfo
  {pages} {1237} (\bibinfo {year} {1983})}\BibitemShut {NoStop}%
\bibitem [{\citenamefont {{von Niessen}}\ \emph {et~al.}(1984)\citenamefont
  {{von Niessen}}, \citenamefont {Schirmer},\ and\ \citenamefont
  {Cederbaum}}]{Nie84}%
  \BibitemOpen
  \bibfield  {author} {\bibinfo {author} {\bibfnamefont {W.}~\bibnamefont {{von
  Niessen}}}, \bibinfo {author} {\bibfnamefont {J.}~\bibnamefont {Schirmer}},\
  and\ \bibinfo {author} {\bibfnamefont {L.}~\bibnamefont {Cederbaum}},\
  }\bibfield  {title} {\bibinfo {title} {{Computational methods for the
  one-particle green's function}},\ }\href
  {https://doi.org/https://doi.org/10.1016/0167-7977(84)90002-9} {\bibfield
  {journal} {\bibinfo  {journal} {Computer Physics Reports}\ }\textbf {\bibinfo
  {volume} {1}},\ \bibinfo {pages} {57} (\bibinfo {year} {1984})}\BibitemShut
  {NoStop}%
\bibitem [{\citenamefont {Demartini}\ \emph {et~al.}(2026)\citenamefont
  {Demartini}, \citenamefont {Berger}, \citenamefont {Blanchon}, \citenamefont
  {Duguet}, \citenamefont {Lacroix}, \citenamefont {Romaniello},\ and\
  \citenamefont {Som\`a}}]{Demartini_2026}%
  \BibitemOpen
  \bibfield  {author} {\bibinfo {author} {\bibfnamefont {T.}~\bibnamefont
  {Demartini}}, \bibinfo {author} {\bibfnamefont {J.~A.}\ \bibnamefont
  {Berger}}, \bibinfo {author} {\bibfnamefont {G.}~\bibnamefont {Blanchon}},
  \bibinfo {author} {\bibfnamefont {T.}~\bibnamefont {Duguet}}, \bibinfo
  {author} {\bibfnamefont {D.}~\bibnamefont {Lacroix}}, \bibinfo {author}
  {\bibfnamefont {P.}~\bibnamefont {Romaniello}},\ and\ \bibinfo {author}
  {\bibfnamefont {V.}~\bibnamefont {Som\`a}},\ }\href
  {https://arxiv.org/abs/2607.26609} {\bibinfo {title} {{Algebraic Diagrammatic
  Construction of the Multichannel Dyson Equation}}} (\bibinfo {year} {2026}),\
  \Eprint {https://arxiv.org/abs/2607.26609} {arXiv:2607.26609} \BibitemShut
  {NoStop}%
\bibitem [{\citenamefont {Schirmer}\ and\ \citenamefont
  {Cederbaum}(1978)}]{Sch78}%
  \BibitemOpen
  \bibfield  {author} {\bibinfo {author} {\bibfnamefont {J.}~\bibnamefont
  {Schirmer}}\ and\ \bibinfo {author} {\bibfnamefont {L.~S.}\ \bibnamefont
  {Cederbaum}},\ }\bibfield  {title} {\bibinfo {title} {{The two-particle-hole
  Tamm-Dancoff approximation (2ph-TDA) equations for closed-shell atoms and
  molecules}},\ }\href {https://doi.org/10.1088/0022-3700/11/11/006} {\bibfield
   {journal} {\bibinfo  {journal} {Journal of Physics B: Atomic and Molecular
  Physics}\ }\textbf {\bibinfo {volume} {11}},\ \bibinfo {pages} {1889}
  (\bibinfo {year} {1978})}\BibitemShut {NoStop}%
\bibitem [{\citenamefont {Keizer}\ \emph {et~al.}(2026)\citenamefont {Keizer},
  \citenamefont {Paggi}, \citenamefont {Berger}, \citenamefont {Romaniello},\
  and\ \citenamefont {F\"orster}}]{Keizer_2026}%
  \BibitemOpen
  \bibfield  {author} {\bibinfo {author} {\bibfnamefont {M.}~\bibnamefont
  {Keizer}}, \bibinfo {author} {\bibfnamefont {S.}~\bibnamefont {Paggi}},
  \bibinfo {author} {\bibfnamefont {J.~A.}\ \bibnamefont {Berger}}, \bibinfo
  {author} {\bibfnamefont {P.}~\bibnamefont {Romaniello}},\ and\ \bibinfo
  {author} {\bibfnamefont {A.}~\bibnamefont {F\"orster}},\ }\href
  {https://arxiv.org/abs/2608.25669} {\bibinfo {title} {{Benchmark of
  Multi-Channel Dyson Equation and Algebraic Diagrammatic Construction Methods
  for molecules}}} (\bibinfo {year} {2026}),\ \Eprint
  {https://arxiv.org/abs/2608.25669} {arXiv:2608.25669} \BibitemShut {NoStop}%
\bibitem [{\citenamefont {Hirofumi}\ \emph {et~al.}(2018)\citenamefont
  {Hirofumi}, \citenamefont {Kosugi}, \citenamefont {Furukawa},\ and\
  \citenamefont {Matsushita}}]{Nishi_2018}%
  \BibitemOpen
  \bibfield  {author} {\bibinfo {author} {\bibfnamefont {N.}~\bibnamefont
  {Hirofumi}}, \bibinfo {author} {\bibfnamefont {T.}~\bibnamefont {Kosugi}},
  \bibinfo {author} {\bibfnamefont {Y.}~\bibnamefont {Furukawa}},\ and\
  \bibinfo {author} {\bibfnamefont {Y.-i.}\ \bibnamefont {Matsushita}},\
  }\bibfield  {title} {\bibinfo {title} {Quasiparticle energy spectra of
  isolated atoms from coupled-cluster singles and doubles (ccsd): Comparison
  with exact ci calculations},\ }\href@noop {} {\bibfield  {journal} {\bibinfo
  {journal} {J. Chem. Phys.}\ }\textbf {\bibinfo {volume} {149}} (\bibinfo
  {year} {2018})}\BibitemShut {NoStop}%
\bibitem [{\citenamefont {Haydock}\ \emph {et~al.}(1972)\citenamefont
  {Haydock}, \citenamefont {Heine},\ and\ \citenamefont {Kelly}}]{Hay72}%
  \BibitemOpen
  \bibfield  {author} {\bibinfo {author} {\bibfnamefont {R.}~\bibnamefont
  {Haydock}}, \bibinfo {author} {\bibfnamefont {V.}~\bibnamefont {Heine}},\
  and\ \bibinfo {author} {\bibfnamefont {M.~J.}\ \bibnamefont {Kelly}},\
  }\bibfield  {title} {\bibinfo {title} {Electronic structure based on the
  local atomic environment for tight-binding bands},\ }\href
  {https://doi.org/10.1088/0022-3719/5/20/004} {\bibfield  {journal} {\bibinfo
  {journal} {Journal of Physics C: Solid State Physics}\ }\textbf {\bibinfo
  {volume} {5}},\ \bibinfo {pages} {2845} (\bibinfo {year} {1972})}\BibitemShut
  {NoStop}%
\bibitem [{\citenamefont {Sun}\ \emph {et~al.}(2020)\citenamefont {Sun},
  \citenamefont {Zhang}, \citenamefont {Banerjee}, \citenamefont {Bao},
  \citenamefont {Barbry}, \citenamefont {Blunt}, \citenamefont {Bogdanov},
  \citenamefont {Booth}, \citenamefont {Chen}, \citenamefont {Cui},
  \citenamefont {Eriksen}, \citenamefont {Gao}, \citenamefont {Guo},
  \citenamefont {Hermann}, \citenamefont {Hermes}, \citenamefont {Koh},
  \citenamefont {Koval}, \citenamefont {Lehtola}, \citenamefont {Li},
  \citenamefont {Liu}, \citenamefont {Mardirossian}, \citenamefont {McClain},
  \citenamefont {Motta}, \citenamefont {Mussard}, \citenamefont {Pham},
  \citenamefont {Pulkin}, \citenamefont {Purwanto}, \citenamefont {Robinson},
  \citenamefont {Ronca}, \citenamefont {Sayfutyarova}, \citenamefont
  {Scheurer}, \citenamefont {Schurkus}, \citenamefont {Smith}, \citenamefont
  {Sun}, \citenamefont {Sun}, \citenamefont {Upadhyay}, \citenamefont {Wagner},
  \citenamefont {Wang}, \citenamefont {White}, \citenamefont {Whitfield},
  \citenamefont {Williamson}, \citenamefont {Wouters}, \citenamefont {Yang},
  \citenamefont {Yu}, \citenamefont {Zhu}, \citenamefont {Berkelbach},
  \citenamefont {Sharma}, \citenamefont {Sokolov},\ and\ \citenamefont
  {Chan}}]{Sun_2020}%
  \BibitemOpen
  \bibfield  {author} {\bibinfo {author} {\bibfnamefont {Q.}~\bibnamefont
  {Sun}}, \bibinfo {author} {\bibfnamefont {X.}~\bibnamefont {Zhang}}, \bibinfo
  {author} {\bibfnamefont {S.}~\bibnamefont {Banerjee}}, \bibinfo {author}
  {\bibfnamefont {P.}~\bibnamefont {Bao}}, \bibinfo {author} {\bibfnamefont
  {M.}~\bibnamefont {Barbry}}, \bibinfo {author} {\bibfnamefont {N.~S.}\
  \bibnamefont {Blunt}}, \bibinfo {author} {\bibfnamefont {N.~A.}\ \bibnamefont
  {Bogdanov}}, \bibinfo {author} {\bibfnamefont {G.~H.}\ \bibnamefont {Booth}},
  \bibinfo {author} {\bibfnamefont {J.}~\bibnamefont {Chen}}, \bibinfo {author}
  {\bibfnamefont {Z.-H.}\ \bibnamefont {Cui}}, \bibinfo {author} {\bibfnamefont
  {J.~J.}\ \bibnamefont {Eriksen}}, \bibinfo {author} {\bibfnamefont
  {Y.}~\bibnamefont {Gao}}, \bibinfo {author} {\bibfnamefont {S.}~\bibnamefont
  {Guo}}, \bibinfo {author} {\bibfnamefont {J.}~\bibnamefont {Hermann}},
  \bibinfo {author} {\bibfnamefont {M.~R.}\ \bibnamefont {Hermes}}, \bibinfo
  {author} {\bibfnamefont {K.}~\bibnamefont {Koh}}, \bibinfo {author}
  {\bibfnamefont {P.}~\bibnamefont {Koval}}, \bibinfo {author} {\bibfnamefont
  {S.}~\bibnamefont {Lehtola}}, \bibinfo {author} {\bibfnamefont
  {Z.}~\bibnamefont {Li}}, \bibinfo {author} {\bibfnamefont {J.}~\bibnamefont
  {Liu}}, \bibinfo {author} {\bibfnamefont {N.}~\bibnamefont {Mardirossian}},
  \bibinfo {author} {\bibfnamefont {J.~D.}\ \bibnamefont {McClain}}, \bibinfo
  {author} {\bibfnamefont {M.}~\bibnamefont {Motta}}, \bibinfo {author}
  {\bibfnamefont {B.}~\bibnamefont {Mussard}}, \bibinfo {author} {\bibfnamefont
  {H.~Q.}\ \bibnamefont {Pham}}, \bibinfo {author} {\bibfnamefont
  {A.}~\bibnamefont {Pulkin}}, \bibinfo {author} {\bibfnamefont
  {W.}~\bibnamefont {Purwanto}}, \bibinfo {author} {\bibfnamefont {P.~J.}\
  \bibnamefont {Robinson}}, \bibinfo {author} {\bibfnamefont {E.}~\bibnamefont
  {Ronca}}, \bibinfo {author} {\bibfnamefont {E.~R.}\ \bibnamefont
  {Sayfutyarova}}, \bibinfo {author} {\bibfnamefont {M.}~\bibnamefont
  {Scheurer}}, \bibinfo {author} {\bibfnamefont {H.~F.}\ \bibnamefont
  {Schurkus}}, \bibinfo {author} {\bibfnamefont {J.~E.~T.}\ \bibnamefont
  {Smith}}, \bibinfo {author} {\bibfnamefont {C.}~\bibnamefont {Sun}}, \bibinfo
  {author} {\bibfnamefont {S.-N.}\ \bibnamefont {Sun}}, \bibinfo {author}
  {\bibfnamefont {S.}~\bibnamefont {Upadhyay}}, \bibinfo {author}
  {\bibfnamefont {L.~K.}\ \bibnamefont {Wagner}}, \bibinfo {author}
  {\bibfnamefont {X.}~\bibnamefont {Wang}}, \bibinfo {author} {\bibfnamefont
  {A.}~\bibnamefont {White}}, \bibinfo {author} {\bibfnamefont {J.~D.}\
  \bibnamefont {Whitfield}}, \bibinfo {author} {\bibfnamefont {M.~J.}\
  \bibnamefont {Williamson}}, \bibinfo {author} {\bibfnamefont
  {S.}~\bibnamefont {Wouters}}, \bibinfo {author} {\bibfnamefont
  {J.}~\bibnamefont {Yang}}, \bibinfo {author} {\bibfnamefont {J.~M.}\
  \bibnamefont {Yu}}, \bibinfo {author} {\bibfnamefont {T.}~\bibnamefont
  {Zhu}}, \bibinfo {author} {\bibfnamefont {T.~C.}\ \bibnamefont {Berkelbach}},
  \bibinfo {author} {\bibfnamefont {S.}~\bibnamefont {Sharma}}, \bibinfo
  {author} {\bibfnamefont {A.~Y.}\ \bibnamefont {Sokolov}},\ and\ \bibinfo
  {author} {\bibfnamefont {G.~K.-L.}\ \bibnamefont {Chan}},\ }\bibfield
  {title} {\bibinfo {title} {Recent developments in the pyscf program
  package},\ }\href {https://doi.org/10.1063/5.0006074} {\bibfield  {journal}
  {\bibinfo  {journal} {The Journal of Chemical Physics}\ }\textbf {\bibinfo
  {volume} {153}},\ \bibinfo {pages} {024109} (\bibinfo {year}
  {2020})}\BibitemShut {NoStop}%
\bibitem [{\citenamefont {Bruneval}\ \emph {et~al.}(2016)\citenamefont
  {Bruneval}, \citenamefont {Rangel}, \citenamefont {Hamed}, \citenamefont
  {Shao}, \citenamefont {Yang},\ and\ \citenamefont {Neaton}}]{Bruneval_2016}%
  \BibitemOpen
  \bibfield  {author} {\bibinfo {author} {\bibfnamefont {F.}~\bibnamefont
  {Bruneval}}, \bibinfo {author} {\bibfnamefont {T.}~\bibnamefont {Rangel}},
  \bibinfo {author} {\bibfnamefont {S.~M.}\ \bibnamefont {Hamed}}, \bibinfo
  {author} {\bibfnamefont {M.}~\bibnamefont {Shao}}, \bibinfo {author}
  {\bibfnamefont {C.}~\bibnamefont {Yang}},\ and\ \bibinfo {author}
  {\bibfnamefont {J.~B.}\ \bibnamefont {Neaton}},\ }\bibfield  {title}
  {\bibinfo {title} {molgw 1: Many-body perturbation theory software for atoms,
  molecules, and clusters},\ }\href
  {https://doi.org/https://doi.org/10.1016/j.cpc.2016.06.019} {\bibfield
  {journal} {\bibinfo  {journal} {Computer Physics Communications}\ }\textbf
  {\bibinfo {volume} {208}},\ \bibinfo {pages} {149} (\bibinfo {year}
  {2016})}\BibitemShut {NoStop}%
\bibitem [{\citenamefont {Marie}\ \emph {et~al.}(2026)\citenamefont {Marie},
  \citenamefont {Burth},\ and\ \citenamefont {Loos}}]{Marie_2026_rel}%
  \BibitemOpen
  \bibfield  {author} {\bibinfo {author} {\bibfnamefont {A.}~\bibnamefont
  {Marie}}, \bibinfo {author} {\bibfnamefont {L.}~\bibnamefont {Burth}},\ and\
  \bibinfo {author} {\bibfnamefont {P.-F.}\ \bibnamefont {Loos}},\ }\bibfield
  {title} {\bibinfo {title} {Reference energies for non-relativistic core
  ionization potentials},\ }\href {https://doi.org/10.1021/acs.jctc.6c00713}
  {\bibfield  {journal} {\bibinfo  {journal} {Journal of Chemical Theory and
  Computation}\ }\textbf {\bibinfo {volume} {22}},\ \bibinfo {pages} {6609}
  (\bibinfo {year} {2026})}\BibitemShut {NoStop}%
\bibitem [{\citenamefont {Deilmann}\ \emph {et~al.}(2016)\citenamefont
  {Deilmann}, \citenamefont {Dr\"uppel},\ and\ \citenamefont
  {Rohlfing}}]{Deilmann_2016}%
  \BibitemOpen
  \bibfield  {author} {\bibinfo {author} {\bibfnamefont {T.}~\bibnamefont
  {Deilmann}}, \bibinfo {author} {\bibfnamefont {M.}~\bibnamefont
  {Dr\"uppel}},\ and\ \bibinfo {author} {\bibfnamefont {M.}~\bibnamefont
  {Rohlfing}},\ }\bibfield  {title} {\bibinfo {title} {Three-particle
  correlation from a many-body perspective: Trions in a carbon nanotube},\
  }\href {https://doi.org/10.1103/PhysRevLett.116.196804} {\bibfield  {journal}
  {\bibinfo  {journal} {Phys. Rev. Lett.}\ }\textbf {\bibinfo {volume} {116}},\
  \bibinfo {pages} {196804} (\bibinfo {year} {2016})}\BibitemShut {NoStop}%
\bibitem [{\citenamefont {Torche}\ and\ \citenamefont
  {Bester}(2019)}]{Torche_2019}%
  \BibitemOpen
  \bibfield  {author} {\bibinfo {author} {\bibfnamefont {A.}~\bibnamefont
  {Torche}}\ and\ \bibinfo {author} {\bibfnamefont {G.}~\bibnamefont
  {Bester}},\ }\bibfield  {title} {\bibinfo {title} {First-principles many-body
  theory for charged and neutral excitations: Trion fine structure splitting in
  transition metal dichalcogenides},\ }\href
  {https://doi.org/10.1103/PhysRevB.100.201403} {\bibfield  {journal} {\bibinfo
   {journal} {Phys. Rev. B}\ }\textbf {\bibinfo {volume} {100}},\ \bibinfo
  {pages} {201403} (\bibinfo {year} {2019})}\BibitemShut {NoStop}%
\bibitem [{\citenamefont {Di~Sabatino}\ \emph {et~al.}(2022)\citenamefont
  {Di~Sabatino}, \citenamefont {Koskelo}, \citenamefont {Berger},\ and\
  \citenamefont {Romaniello}}]{Sab22}%
  \BibitemOpen
  \bibfield  {author} {\bibinfo {author} {\bibfnamefont {S.}~\bibnamefont
  {Di~Sabatino}}, \bibinfo {author} {\bibfnamefont {J.}~\bibnamefont
  {Koskelo}}, \bibinfo {author} {\bibfnamefont {J.~A.}\ \bibnamefont
  {Berger}},\ and\ \bibinfo {author} {\bibfnamefont {P.}~\bibnamefont
  {Romaniello}},\ }\bibfield  {title} {\bibinfo {title} {Introducing screening
  in one-body density matrix functionals: Impact on charged excitations of
  model systems via the extended koopmans' theorem},\ }\href
  {https://doi.org/10.1103/PhysRevB.105.235123} {\bibfield  {journal} {\bibinfo
   {journal} {Phys. Rev. B}\ }\textbf {\bibinfo {volume} {105}},\ \bibinfo
  {pages} {235123} (\bibinfo {year} {2022})}\BibitemShut {NoStop}%
\bibitem [{\citenamefont {Di~Sabatino}\ \emph {et~al.}(2023)\citenamefont
  {Di~Sabatino}, \citenamefont {Koskelo}, \citenamefont {Berger},\ and\
  \citenamefont {Romaniello}}]{Dis23-1}%
  \BibitemOpen
  \bibfield  {author} {\bibinfo {author} {\bibfnamefont {S.}~\bibnamefont
  {Di~Sabatino}}, \bibinfo {author} {\bibfnamefont {J.}~\bibnamefont
  {Koskelo}}, \bibinfo {author} {\bibfnamefont {J.~A.}\ \bibnamefont
  {Berger}},\ and\ \bibinfo {author} {\bibfnamefont {P.}~\bibnamefont
  {Romaniello}},\ }\bibfield  {title} {\bibinfo {title} {Screened extended
  koopmans' theorem: Photoemission at weak and strong correlation},\ }\href
  {https://doi.org/10.1103/PhysRevB.107.035111} {\bibfield  {journal} {\bibinfo
   {journal} {Phys. Rev. B}\ }\textbf {\bibinfo {volume} {107}},\ \bibinfo
  {pages} {035111} (\bibinfo {year} {2023})}\BibitemShut {NoStop}%
\bibitem [{\citenamefont {Venkatareddy}\ \emph {et~al.}(2025)\citenamefont
  {Venkatareddy}, \citenamefont {Ghosh}, \citenamefont {Krishnamurthy},\ and\
  \citenamefont {Jain}}]{Venkatareddy_2025}%
  \BibitemOpen
  \bibfield  {author} {\bibinfo {author} {\bibfnamefont {N.}~\bibnamefont
  {Venkatareddy}}, \bibinfo {author} {\bibfnamefont {V.}~\bibnamefont {Ghosh}},
  \bibinfo {author} {\bibfnamefont {H.~R.}\ \bibnamefont {Krishnamurthy}},\
  and\ \bibinfo {author} {\bibfnamefont {M.}~\bibnamefont {Jain}},\ }\bibfield
  {title} {\bibinfo {title} {Double excitations in molecules using screened
  configuration interaction},\ }\href
  {https://doi.org/10.1021/acs.jctc.5c01484} {\bibfield  {journal} {\bibinfo
  {journal} {Journal of Chemical Theory and Computation}\ }\textbf {\bibinfo
  {volume} {21}},\ \bibinfo {pages} {12162} (\bibinfo {year} {2025})},\ \Eprint
  {https://arxiv.org/abs/https://doi.org/10.1021/acs.jctc.5c01484}
  {https://doi.org/10.1021/acs.jctc.5c01484} \BibitemShut {NoStop}%
\bibitem [{\citenamefont {van Setten}\ \emph {et~al.}(2013)\citenamefont {van
  Setten}, \citenamefont {Weigend},\ and\ \citenamefont {Evers}}]{vanSet13}%
  \BibitemOpen
  \bibfield  {author} {\bibinfo {author} {\bibfnamefont {M.~J.}\ \bibnamefont
  {van Setten}}, \bibinfo {author} {\bibfnamefont {F.}~\bibnamefont
  {Weigend}},\ and\ \bibinfo {author} {\bibfnamefont {F.}~\bibnamefont
  {Evers}},\ }\bibfield  {title} {\bibinfo {title} {The gw-method for quantum
  chemistry applications: Theory and implementation},\ }\href
  {https://doi.org/10.1021/ct300648t} {\bibfield  {journal} {\bibinfo
  {journal} {Journal of Chemical Theory and Computation}\ }\textbf {\bibinfo
  {volume} {9}},\ \bibinfo {pages} {232} (\bibinfo {year} {2013})}\BibitemShut
  {NoStop}%
\end{thebibliography}
\end{document}